\documentclass[reprint,twocolumn]{revtex4-2}
\usepackage{amsfonts}
\usepackage{amsmath}
\usepackage{amssymb}
\usepackage{charter}
\usepackage{graphicx}
\usepackage{float}
\usepackage{amsmath}
\usepackage{amssymb}
\usepackage{charter}
\usepackage{graphicx}
\usepackage{subfigure}
\usepackage{graphicx}
\usepackage{epstopdf}
\usepackage{color}
\usepackage{tocvsec2}
\usepackage{enumerate}
\usepackage{graphicx}
\usepackage{dcolumn}
\usepackage{bm}

\begin{document}

\title{Phase estimation via photon subtraction at the output of the hybrid
interferometer}
\author{Qisi Zhou$^{1}$, Tao Jiang$^{1}$, Qingqian Kang$^{1,2}$, Teng Zhao$%
^{1,3}$, Xin Su$^{1}$}
\author{Cunjin Liu$^{1\ast }$}
\author{Liyun Hu$^{1}$}
\thanks{Corresponding author. lcjwelldone@126.com; hlyun@jxnu.edu.cn}

\begin{abstract}
The hybrid interferometer integrating an optical parametric amplifier and a
beam splitter has the potential to outperform the SU(1,1) interferometer.
However, photon loss remains a critical limitation for practical
implementation. To address this challenge, we propose a quantum metrology
scheme utilizing multi-photon\ subtraction at the output and replacing the\
conventional 50:50 beam splitter with a variable beam splitter to enhance
robustness against photon loss. We\ employ a coherent state and a vacuum
state as inputs and perform homodyne detection.\ Our results show that the
selection of input modes significantly affects phase estimation, and
optimizing the beam splitter's transmittance is crucial for maximizing phase
sensitivity in lossy conditions. Furthermore, photon subtraction markedly
improves phase sensitivity, quantum Fisher information, and robustness
against noise. Our scheme achieves sensitivities beyond the Heisenberg limit
even under 20\% photon loss.

\end{abstract}

\maketitle

\affiliation{$^{{\small 1}}$\textit{Center for Quantum Science and Technology, Jiangxi
Normal University, Nanchang 330022, China}\\
$^{{\small 2}}$\textit{Department of Physics, Jiangxi Normal University Science and
Technology College, Nanchang 330022, China}\\
$^{{\small 3}}$\textit{Institute for Military-Civilian Integration of Jiangxi Province,
 Nanchang 330200, China}}

\section{Introduction}

Quantum precision measurement, a core area of quantum information science,
demonstrates revolutionary potential in emerging technologies such as
gravitational wave detection \cite{1,2}, quantum imaging \cite{3}, and
atomic clock calibration \cite{4,5}. Optical interferometers serve as key
tools for high-precision phase measurements, where sensitivity directly
determines detection accuracy \cite{6}. Traditional Mach-Zehnder
interferometers (MZIs) using classical light sources are constrained by
vacuum noise, limiting phase sensitivity to the standard quantum limit (SQL) 
\cite{7}. To overcome this limit, many researchers have proposed using
quantum light sources as input, such as squeezed states \cite{8}, Fock
states \cite{9}, NOON states \cite{10}. Under quantum schemes, phase
sensitivity can surpass the SQL and even reach the Heisenberg limit (HL) 
\cite{11,12}. Another approach to surpass the SQL is the SU(1,1)
interferometer proposed by Yurke in 1981, which replaces two beam splitters
(BSs) with optical parametric amplifiers (OPAs) \cite{13,14,15}.\ The
standard setup must satisfy the phase-matching condition $\theta _{2}-\theta
_{1}$ $=\pi $, where $\theta _{1}$ and $\theta _{2}$ are the phases of the
first and second OPA, respectively \cite{aa18}. Studies show that it can
exceed the SQL at low photon numbers \cite{16}. Li et al. showed that using
coherent states and squeezed vacuum states with homodyne detection allows
sensitivity to approach HL \cite{17}.

In practice, the SU(1,1) interferometer faces major challenges, including
sensitivity to photon loss \cite{20,21,22} and photon absorption during the
second nonlinear process \cite{aa19}, which limit the effectiveness of
quantum resources \cite{23,24}. To overcome these limitations, researchers
have proposed various improved configurations, such as unbalanced-gain
SU(1,1) interferometers \cite{26,27}, truncated SU(1,1) interferometers \cite%
{29,30,31}, and atom-light hybrid interferometer \cite{aa32}. Moreover, the
OPA-BS hybrid interferometer has been widely studied due to its structural
advantages \cite{aa19}, preserving non-linear gain properties while reducing
quantum resource consumption and improving photon efficiency \cite{32}. Kong
et al. \cite{33} demonstrate that the OPA-BS structure can cancel quantum
noise and improve the output signal-to-noise ratio.\ Zhang et al. proposed a
hybrid interferometer with two coherent states inputs and homodyne
detection, achieving sub-SQL phase sensitivity and approaching the quantum
Cram\'{e}r-Rao bound (QCRB). Shao et al. showed that under the optimal
conditions, its phase sensitivity can exceed the SQL.\ Although the OPA-BS
hybrid interferometer has made significant progress, it still faces some
challenges, particularly the impact of loss on phase sensitivity \cite{aaa34}%
.

To improve phase sensitivity and system robustness, non-Gaussian\ operations
(such as photon subtraction (PS), photon addition, and photon catalysis)
have been widely applied in phase estimation \cite{34}. For example, Gong et
al. showed that subtracting $m$ photons from a squeezed vacuum state ($m$%
-PS-SVS) significantly enhances the quantum Fisher information (QFI) of an
SU(1,1) interferometer, with sensitivity approaching the HL as more photons
are subtracted \cite{35}. Xu et al. further showed that photon addition
within the interferometer effectively reduces photon loss and improves
sensitivity \cite{21}. However, most studies focus on applying photon
operations at the input or within traditional MZI and SU(1,1)
interferometers. Research on non-Gaussian operations at the output of hybrid
interferometers remains limited \cite{36,37,38,39,40}. Output-port PS can
manipulate the quantum state before detection, thereby reducing photon loss
during transmission and potentially offering a new strategy for enhancing
phase sensitivity.

In this parper, we propose an improved OPA-BS hybrid interferometer scheme
with PS at the output and coherent states inputs. We compare the
improvements in phase sensitivity and QFI for two input conditions: coherent
states applied to mode $a$ or mode $b$. Additionly, we investigate the
influence of the transmittance of the variable BS (vBS) on phase sensitivity
under loss conditions, showing that higher transmittance enhances
sensitivity and robustness in high-loss scenarios.

This paper is organized as follows. Sec. II introduces the model of the
modified hybrid interferometer. Sec. III investigates phase sensitivity
under ideal and lossy conditions. Sec. IV analyzes the impact of the
proposed scheme on QFI and QCRB. The conclusions are presented in Sec. V.

\section{Hybrid interferometer model}

In this section, we introduce the modified hybrid interferometer, which
consists of an OPA, a vBS with transmittance $\tau $, and a linear phase
shifter. Two input schemes are considered: one is an $a$-mode coherent state
and $b$-mode vacuum state, i.e., $\left \vert \alpha \right \rangle
_{a}\otimes \left \vert 0\right \rangle _{b}$, where $\alpha =\left \vert
\alpha \right \vert e^{i\theta _{\alpha }}$. The other case is reversed,
i.e., $\left \vert 0\right \rangle _{a}\otimes \left \vert \beta
\right
\rangle _{b}$, where $\beta =\left \vert \beta \right \vert
e^{i\theta _{\beta }}$. The input states are first injected into a pumped
OPA, undergo an unknown phase shift $\phi $ via a phase shifter $U_{\phi
}=\exp [i\phi a^{\dagger }a] $, and then are recombined by a vBS. Finally,
PS and homodyne detection are performed on mode $a$. Here, the OPA process
can be described by a two-mode squeezing operator $S_{g}=\exp [\xi ^{\ast
}ab-\xi a^{\dag }b^{\dag }]$, where ($a$, $b$) and ($a^{\dagger }$, $%
b^{\dagger }$) are the photon annihilation and creation operators,
respectively, and $\xi =ge^{i\theta }$ is the squeezing parameter, with $g$
and $\theta $ denoting the gain factor and OPA phase respectively. In our
scheme, the input-output field operator transformations of a vBS is given by%
\begin{eqnarray}
B_{v}^{\dagger }aB_{v} &=&\sqrt{\tau }a+i\sqrt{1-\tau }b,  \notag \\
B_{v}^{\dagger }bB_{v} &=&i\sqrt{1-\tau }a+\sqrt{\tau }b,  \label{a1}
\end{eqnarray}%
where, $B_{v}$ respects the operator of vBS with transmittance $\tau $,
where $\tau \in \left[ 0,1\right] $. 
\begin{figure}[tph]
\label{Figure1} \centering \includegraphics[width=0.85\columnwidth]{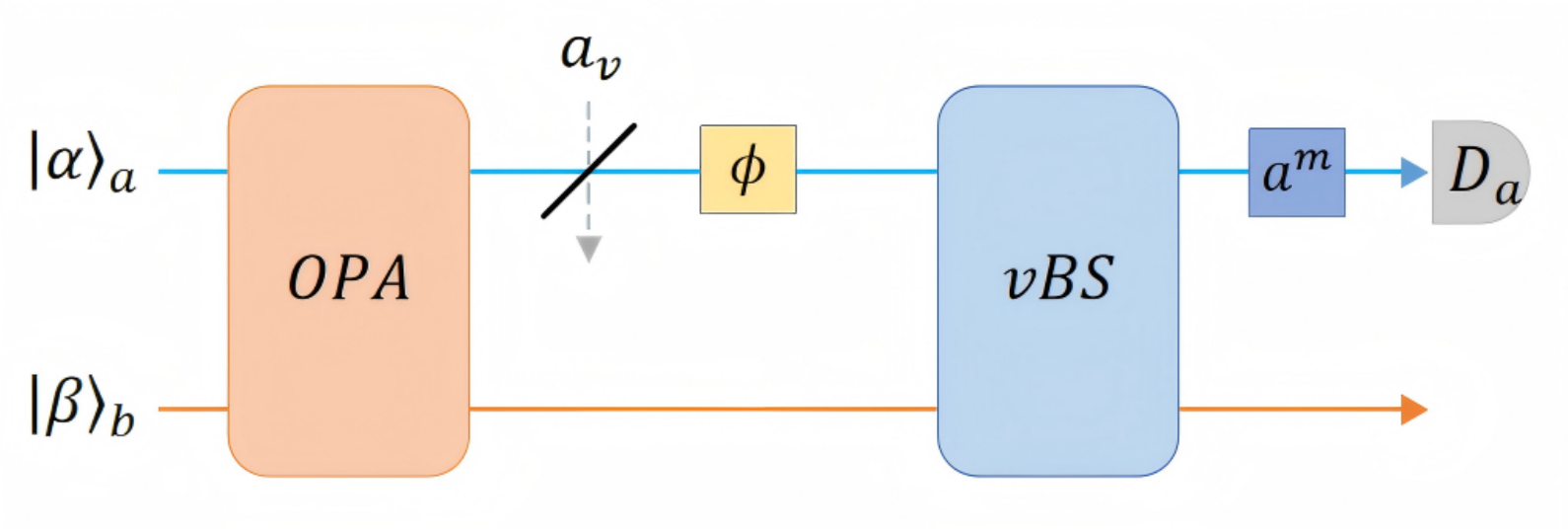}
\caption{Schematic diagram of the hybrid interferometer with PS under photon
loss. Two input configurations are considered: Scheme A $\left( \left \vert 
\protect \psi _{1}\right \rangle _{in}=\left \vert \protect \alpha %
\right
\rangle _{a}\left \vert 0\right \rangle _{b}\right) $ and Scheme B $%
\left( \left \vert \protect \psi _{2}\right \rangle _{in}=\left \vert
0\right
\rangle _{a}\left \vert \protect \beta \right \rangle _{b}\right) $.
OPA denotes an optical parametric amplifier, $\protect \phi $ is the phase
shift, vBS is a variable beam splitter, and $D_{a}$ is the homodyne
detector.\ The\ operator $a^{m}$ corresponds to $m$-photons subtraction, and 
$a_{v}$ denotes vacuum mode.}
\end{figure}

In experimental systems, noise is usually present. Relevant studies have
demonstrated that photon loss within the interferometer exerts a more
significant influence on phase sensitivity \cite{41}. Therefore, we use a
fictitious BS between the OPA and the phase shifter to simulate photon loss
in mode\ $a$, as shown in Fig. 1. This process is described by the following
transformation:%
\begin{equation}
B_{T}^{\dag }aB_{T}=\sqrt{T}a+\sqrt{1-T}a_{v},  \label{a2}
\end{equation}%
where, $B_{T}$ is the operator of the fictitious BS with transmittance\ $T$,
where $T\in \left[ 0,1\right] $, and $a_{v}$ is the vacuum mode. When $T$ $%
=1 $, it indicates no photon loss; when $T=0$, it indicates complete loss.
For simplicity, we set coherent state phases $\theta _{\alpha }=\theta
_{\beta }=0$. In an expanded space, the output state under photon losses
cases can be represented as 
\begin{equation}
\left \vert \Psi \right \rangle _{out}=N_{0}a^{m}B_{v}U_{\phi
}B_{T}S_{g}\left \vert \alpha \right \rangle _{a}\left \vert \beta \right
\rangle _{b}\left \vert 0\right \rangle _{a_{v}},  \label{a3}
\end{equation}%
where $N_{0}$ is the normalization constant, and $m$ is the
photon-subtracted number. When $\alpha \neq 0,\beta =0$, it corresponds to
the output state of Scheme A; when $\alpha =0,\beta \neq 0$, it corresponds
to the output state of Scheme B. The normalization constant can be
calculated as 
\begin{equation}
N_{0}=\frac{1}{\sqrt{D_{m,m}e^{Z_{0}}}},  \label{a4}
\end{equation}%
where%
\begin{equation}
D_{m,m}=\frac{\partial ^{m+m}}{\partial s^{m}\partial t^{m}}\left( \cdot
\right) |_{s=t=0},  \label{a5}
\end{equation}%
and%
\begin{eqnarray}
Z_{0} &=&s^{2}X_{1}+t^{2}X_{1}^{\ast }+sX_{2}\alpha +tX_{2}^{\ast }\alpha 
\notag \\
&&+sX_{3}\beta +tX_{3}^{\ast }\beta +stX_{4}  \label{a6}
\end{eqnarray}%
with%
\begin{eqnarray}
X_{1} &=&-ie^{i\phi }e^{i\theta }\sqrt{\tau T}\sqrt{1-\tau }\sinh g\cosh g, 
\notag \\
X_{2} &=&e^{i\phi }\sqrt{\tau T}\cosh g-ie^{i\theta }\sqrt{1-\tau }\sinh g, 
\notag \\
X_{3} &=&i\sqrt{1-\tau }\cosh g-e^{i\phi }e^{i\theta }\sqrt{\tau T}\sinh g, 
\notag \\
X_{4} &=&\left( 1-\tau +\tau T\right) \sinh ^{2}g.  \label{a7}
\end{eqnarray}%
$m$ is positive integer, while $s$ and $t$ are the differential variables
that become zero after differentiation. Here, we only consider the balanced
parameter configuration with the OPA phase $\theta =\pi $.

\section{The phase sensitivity}

In quantum precision measurement, phase sensitivity depends not only on the
input state and interferometer structure, but also on the detection method.
Common detection methods include intensity detection \cite{42}, homodyne
detection \cite{43}, and parity detection \cite{44}. This section employs
homodyne detection, which extracts the orthogonal components of the optical
field by interfering the signal light with the reference light, at the
output $a$ to study\ phase measurement precision. By the error propagation
equation, the phase sensitivity can be obtained \cite{45} 
\begin{equation}
\Delta \phi =\frac{\sqrt{\left \langle X^{2}\right \rangle -\left \langle
X\right \rangle ^{2}}}{|\partial \left \langle X\right \rangle /\partial
\phi |},  \label{a9}
\end{equation}%
where $X=(a+a^{\dagger })/\sqrt{2}$ is the orthogonal component operator of
the $a$-mode light field, and $\left \langle \cdot \right \rangle =\left.
_{out}\left \langle \Psi \right \vert \cdot \left \vert \Psi \right \rangle
_{out}\right. $. According to Eqs. (\ref{a3}) and (\ref{a9}), the phase
sensitivity can be obtained (see Supplement 1, S1 for details).

\subsection{The phase sensitivity in the ideal case}

Firstly, we investigate the influence of phase $\phi $ and the transmittance
of vBS $\tau $ on phase sensitivity in the ideal case, corresponding to $T=1$%
. 
\begin{figure*}[tbp]
\label{Figure2} {\centering \includegraphics[width=1.75%
\columnwidth]{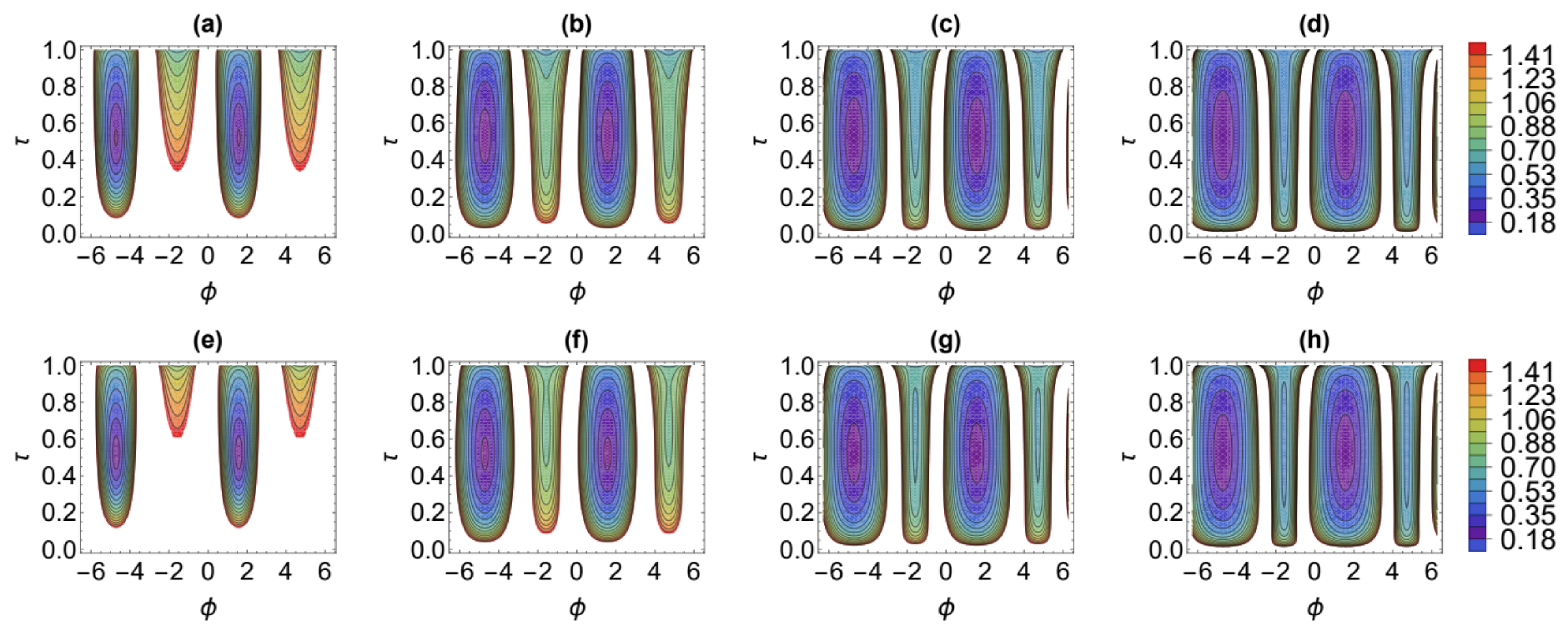}}
\caption{The contour plots of phase sensitivity as a function of parameters $%
\protect \phi $ and $\protect \tau $, fixed with $\protect \alpha (\protect%
\beta )=1$ and $g=1$, where $m$ denotes the PS order. Fig. 2(a)-(d)
correspond respectively to $m=0$, $1$, $2$, and $3$ in Scheme A; Fig.
2(e)-(h)\ correspond to those in Scheme B.}
\end{figure*}
Fig. 2 shows the contour plots of phase sensitivity as a function of
parameters $\phi $ and $\tau $ for two input schemes. Fig. 2(a)-(d)
correspond respectively to $m=0$, $1$, $2$, and $3$ for Scheme A; Fig.\
2(e)-(h)\ correspond to those for Scheme B. The results show that: (i) For
all photon-subtracted number ($m=0,1,2,3$), the optimal phase sensitivity is
achieved when $\phi $ is near $\pi /2$ and $\tau $ is around 0.5,
corresponding to optimal phase matching for this interference as well as the
balanced point of the BS. This parameter setting enables OPA amplification
and PS to synergistically enhance each other. (ii) The behavior is
consistent for both input schemes. Therefore, in the subsequent analysis of
the ideal situation, $\phi $ and $\tau $ will be fixed at $\pi /2$ and 0.5,
respectively, to evaluate the effect of PS on phase estimation. The impact
of photon loss will be discussed separately. 
\begin{figure}[tbp]
\label{Figure3} {\centering \includegraphics[width=0.85%
\columnwidth]{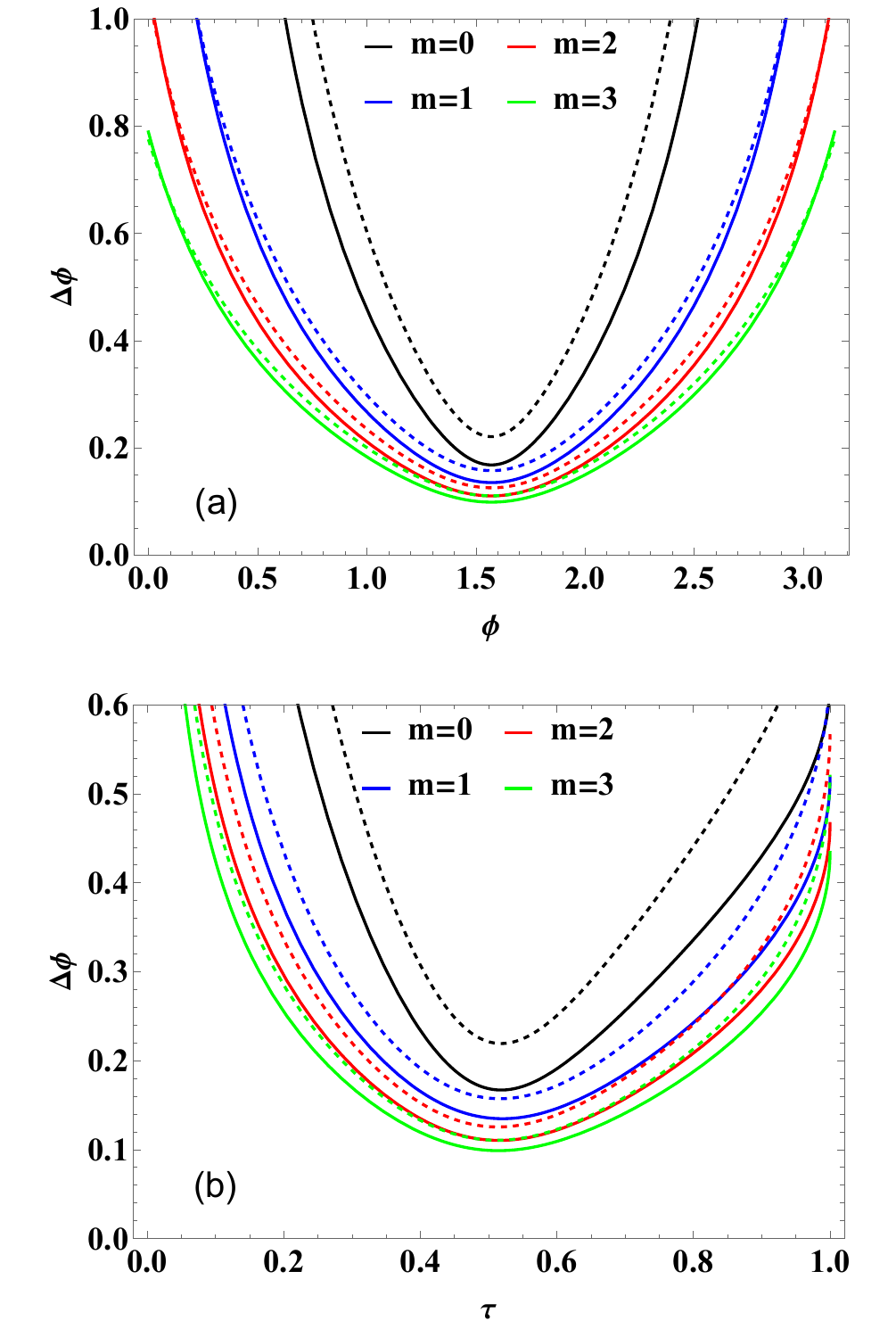}}
\caption{The phase sensitivity as a function of (a) phase $\protect \phi $,
with $\protect \tau =0.5$, $\protect \alpha \left( \protect \beta \right) =1$,
and $g=1$; (b) transmittance $\protect \tau $, with $\protect \phi =\protect%
\pi /2$, $\protect \alpha \left( \protect \beta \right) =1$, and $g=1$. The
solid and dashed lines represent Schemes A and B, respectively.}
\end{figure}

In Fig. 3(a), the phase sensitivity is plotted as a function of the phase $%
\phi $ with fixed $\tau =0.5$. Here, $m$ denotes the photon-subtracted
number, while the solid and dashed lines represent Scheme A and B,
respectively. The following conclusions can be drawn: (i) With identical
parameters, Scheme A outperforms Scheme B; (ii) Phase sensitivity improves
as $m$ increases, and the improvement is more significant when $m$ is small;
(iii) For both schemes, the optimal phase points are near $\pi /2$, and
phase sensitivity declines as the deviation from this point increases, which
is consistent with the previous conclusion; (iv) When the phase deviates
significantly from the optimal value, the enhancement effects of PS for the
two schemes become similar. Fig. 3(b) shows that under a fixed phase $\phi
=\pi /2$, the phase sensitivity first increases and then decreases with $%
\tau $, reaching the optimal value at $\tau =0.5$. Moreover, Scheme A
consistently performs better than Scheme B. In conclusion, when $\tau =0.5$,
the ideal system achieves optimal parameter matching for quantum
interference, which is critical for high-precision phase measurement. 
\begin{figure}[tph]
\label{Figure4} \centering \includegraphics[width=0.85\columnwidth]{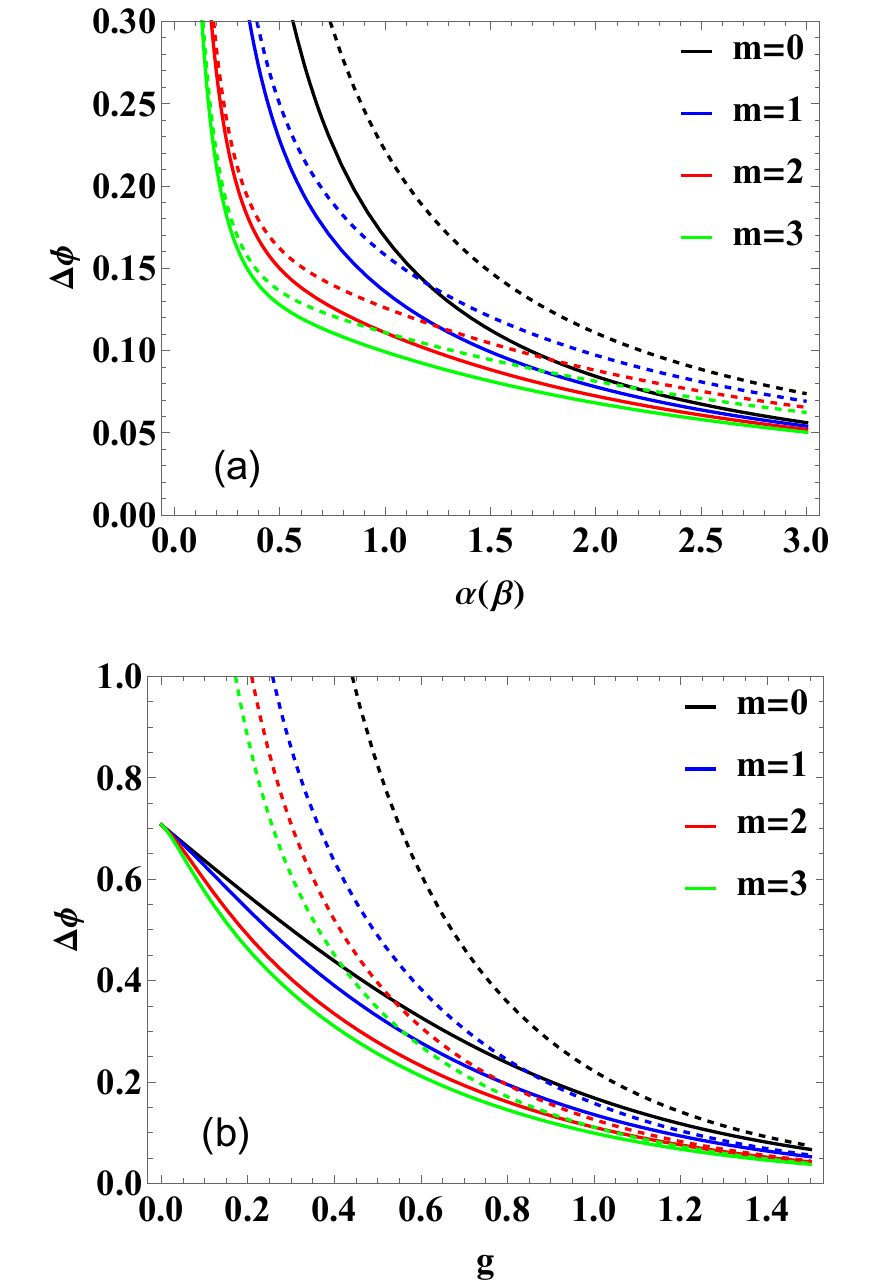}
\caption{The phase sensitivity of PS based on homodyne detection as a
function of (a) $\protect \alpha \left( \protect \beta \right) $ with $\protect%
\phi =\protect \pi /2$, $\protect \tau =0.5$, and $g=1$; (b) $g$ with $\protect%
\phi =\protect \pi /2$, $\protect \tau =0.5$, and $\protect \alpha \left( 
\protect \beta \right) =1$. The solid and dashed lines represent Schemes A
and B, respectively.}
\end{figure}

Next, to investigate the influence of other parameters, we plot the phase
sensitivity as a function of coherent amplitude $\alpha \left( \beta \right) 
$ and gain factor\ $g$,\ in Fig. 4. The solid and dashed lines represent
Scheme A and Scheme B, respectively.\ The results show that: (i) Phase
sensitivity improves with increasing $\alpha \left( \beta \right) $ and $g$.
This is because larger $\alpha \left( \beta \right) $ increases input photon
number, while higher $g$ enhances the nonlinear gain; (ii) As $\alpha \left(
\beta \right) $ increases, the phase sensitivity's improvement decreases; as 
$g$ increases, it first rises and then falls; (iii) Scheme A outperforms
Scheme B, especially at large $\alpha \left( \beta \right) $ and small $g$,
but the difference is not significant when $\alpha \left( \beta \right) $ is
small and $g$ is large. This indicates that input mode choice affects
measurement accuracy. Notably, at large $\alpha \left( \beta \right) $,
Scheme A without PS outperforms Scheme B with PS. When $g\rightarrow 0$,
Scheme A performs well while Scheme B does not. \ 

\subsection{The phase sensitivity in the lossy case}

In practical applications, the system is inevitably affected by photon loss.
To further analyze this impact, we investigate the performance of phase
sensitivity under loss conditions. It is important to note that when loss is
present, the optimal transmittance of vBS may deviate from that of the ideal
case, i.e., 0.5. 
\begin{figure}[tbp]
\label{Figure5} {\centering \includegraphics[width=0.85%
\columnwidth]{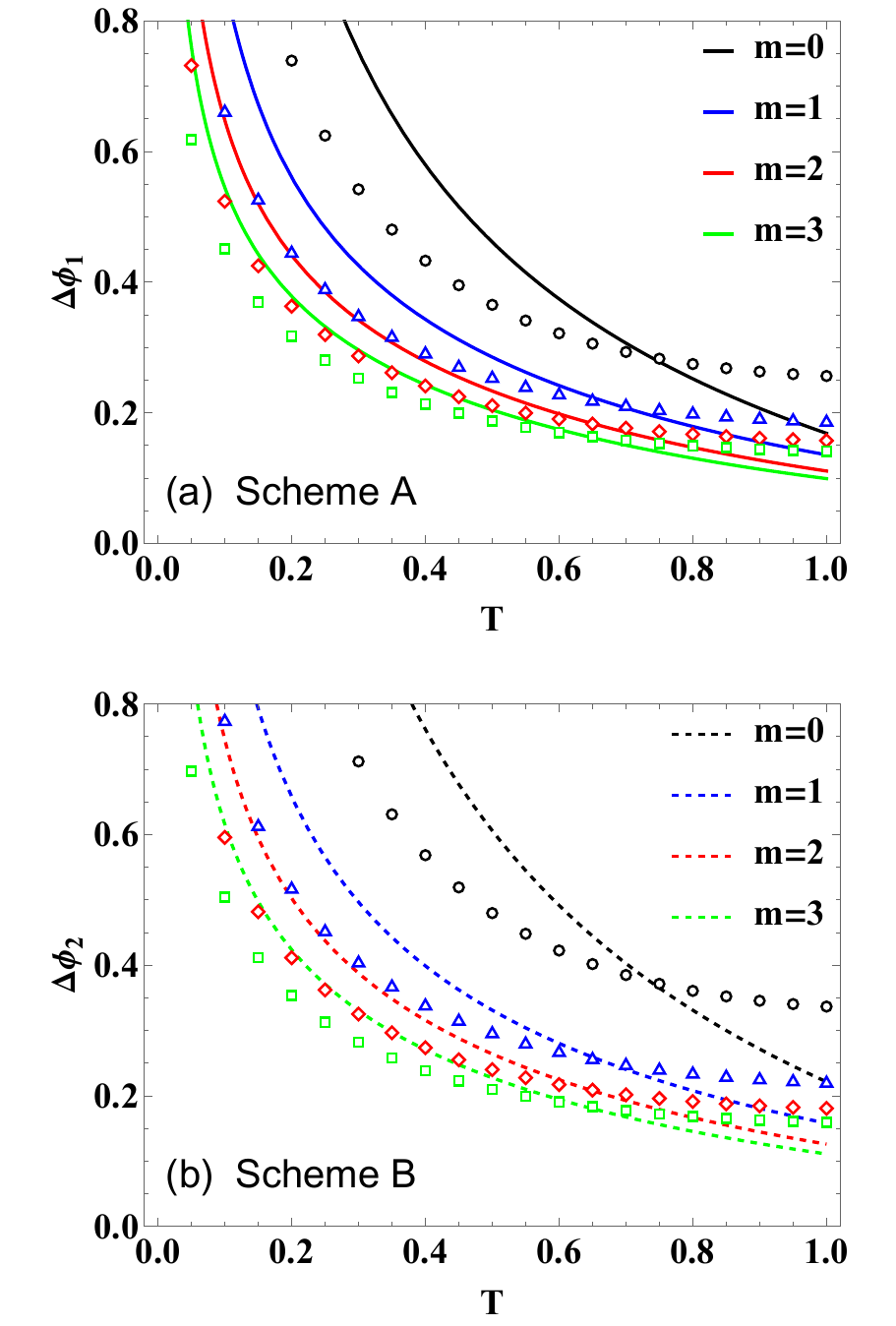}}
\caption{The phase sensitivity of PS as a function of $T$, with $\protect%
\phi =\protect \pi /2$, $\protect \alpha \left( \protect \beta \right) =1$, and 
$g=1$. (a) Scheme A; (b) Scheme B. The solid and dashed line correspond to $%
\protect \tau =0.5$, while the point set corresponds to $\protect \tau =0.7$.}
\end{figure}

Fig. 5 presents the phase sensitivity as a function of transmittance $T$ for
Scheme A (Fig. 5(a)) and Scheme B (Fig. 5(b)), comparing the results for $%
\tau $ fixed at 0.5 and 0.7. Higher transmittance $\tau $ results in more $a$%
-mode light carrying phase shift information being transmitted through the
vBS to the output port. Here, the line shows sensitivity at $\tau $ $=0.5$,
and the point set show it at $\tau =0.7$. It can be seen that: (i) The phase
sensitivity decreases as $T$ decreases; (ii) With fixed parameters, PS
enhances the phase sensitivity\ and improves the system's robustness against
photon loss, especially as $m$ increases; (iii) At low photon loss levels,
the system with $\tau =0.5$ exhibits better sensitivity; however, when loss
exceeds 30\%, the system with $\tau =0.7$ performs better. This indicates
that tuning the transmittance of vBS effectively enhances system performance
under high-loss conditions; (iv) Under loss conditions, Scheme A
consistently outperforms Scheme B, exhibiting higher phase sensitivity and
greater robustness. 
\begin{figure*}[tbp]
\label{Figure6} {\centering \includegraphics[width=1.8\columnwidth]{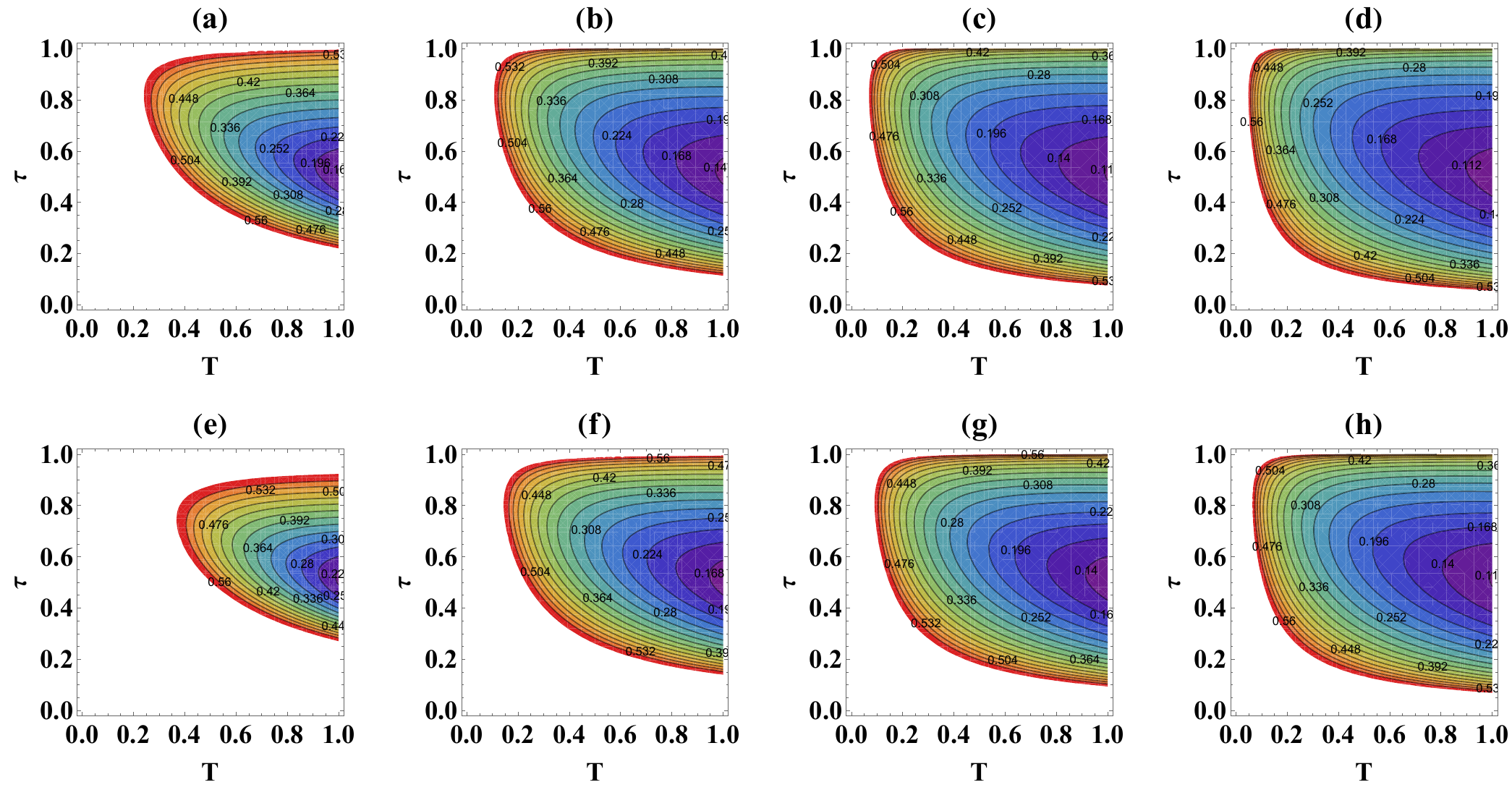}%
}
\caption{The contour plots of phase sensitivity as a function of $T$ and $%
\protect \tau $, with $\protect \phi =\protect \pi /2$, $\protect \alpha \left( 
\protect \beta \right) =1$, $g=1$. The first and second rows represent the
density plots for Scheme A and B, respectively. Figs. (a)-(d) correspond
respectively to $m=0$, $1$, $2$, and $3$ for Scheme A; Figs.\ (e)-(h)\
correspond to those for Scheme B.}
\end{figure*}

To visually analyze\ how vBS's transmittance affects phase sensitivity under
different loss conditions, we present contour plots of phase sensitivity as
a function of $T$ and $\tau $, in Fig. 6. The first and second rows
represent the contour plots for Scheme A and B, respectively, while the
first, second, third, and fourth columns correspond to the photon-subtracted
number $m$ $=0$, $1$, $2$, and $3$, respectively. It can be clearly seen
that: (i) As $m$ increases, the phase sensitivity is enhanced over a broader
parameter range, highlighting the advantage of non-Gaussian operations in
quantum metrology; (ii) Under nearly ideal conditions, the optimal $\tau $
is approximately 0.5. However, as photon loss increases, the optimal $\tau $
shifts away from 0.5, with higher loss corresponding to higher optimal $\tau 
$ values. This demonstrates that the synergy between PS and vBS enhances the
phase sensitivity of our scheme under photon loss conditions, offering
notable robustness and practical potential.

\section{The quantum Fisher information}

The QFI is a key theoretical metric used to assess measurement precision in
the field of quantum metrology \cite{46}. It quantifies the sensitivity of a
quantum state to infinitesimal phase shifts, with a higher QFI indicating a
greater capacity to extract information about the measured parameter.
Improving the QFI through quantum techniques can enhance the phase
measurement sensitivity.

\subsection{QFI in the ideal case}

For a pure state system, the QFI under ideal conditions is given by \cite%
{47,48} (see Supplement 1, S2 for the detailed derivation)%
\begin{equation}
F=4\left[ \left \langle \Psi _{\phi }^{\prime }|\Psi _{\phi }^{\prime
}\right \rangle -\left \vert \left \langle \Psi _{\phi }^{\prime }|\Psi
_{\phi }\right \rangle \right \vert ^{2}\right] ,  \label{b1}
\end{equation}%
where $\left \vert \Psi _{\phi }\right \rangle $ represents the quantum
state after the phase shifter with $\left \vert \Psi _{\phi }^{\prime
}\right \rangle =\partial \left \vert \Psi _{\phi }\right \rangle /\partial
\phi $. To study the effect of our scheme on QFI in the ideal case, we
construct an equivalent theoretical model (see Fig. 7). In this model, $%
U_{P} $ denotes the non-local operation that incorporates the PS operation
at the output port and satisfies $U_{P}=B_{v}^{\dag }a^{m}B_{v}$. We
consider two kinds of input schemes. For convenience, we uniformly represent
the state as $\left \vert \Psi _{\phi }\right \rangle =N_{1}U_{P}U_{\phi
}S_{g}\left \vert \alpha \right \rangle _{a}\left \vert \beta \right \rangle
_{b}$. 
\begin{figure}[tph]
\label{Figure7} \centering \includegraphics[width=0.8\columnwidth]{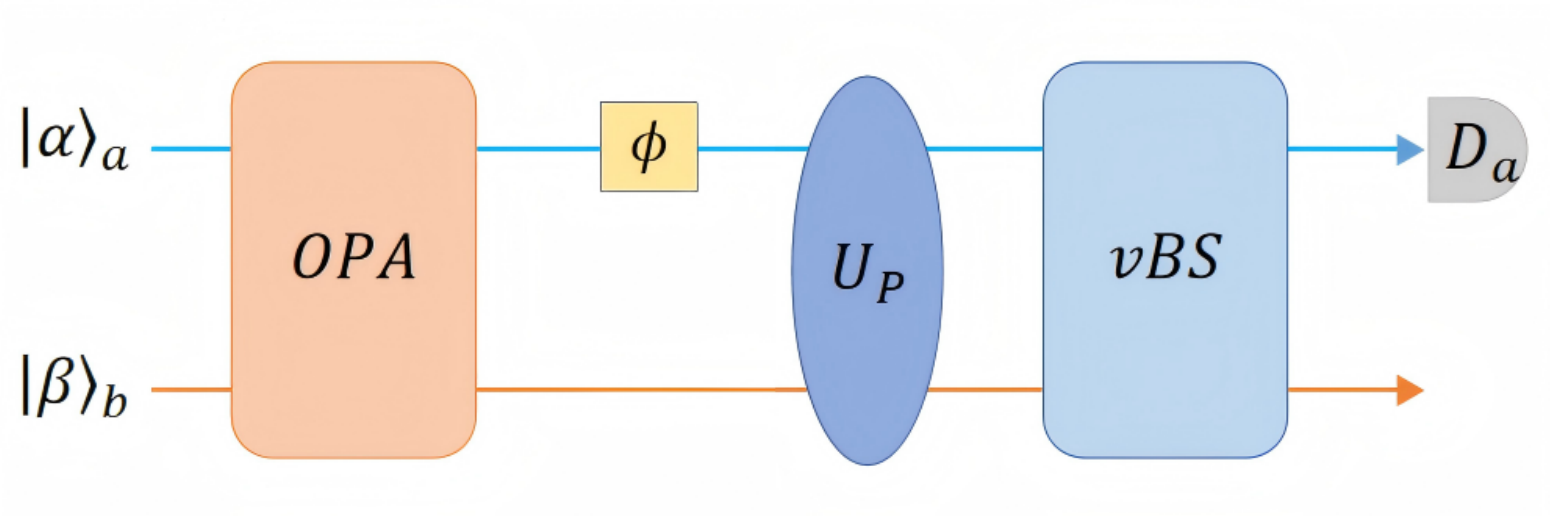}
\caption{Schematic diagram of the equivalent model of the hybrid
interferometer, where $U_{P}$ represents a non-local operation, and $%
U_{P}=B_{v}^{\dag }a^{m}B_{v}$.}
\end{figure}

The QCRB is given by \cite{49}%
\begin{equation}
\Delta \phi _{QCRB}=\frac{1}{\sqrt{vF}},  \label{b2}
\end{equation}%
where $v$ is the number of repeated measurements, typically set to $1$. 
\begin{figure*}[tbp]
\label{Figure8} {\centering \includegraphics[width=1.4\columnwidth]{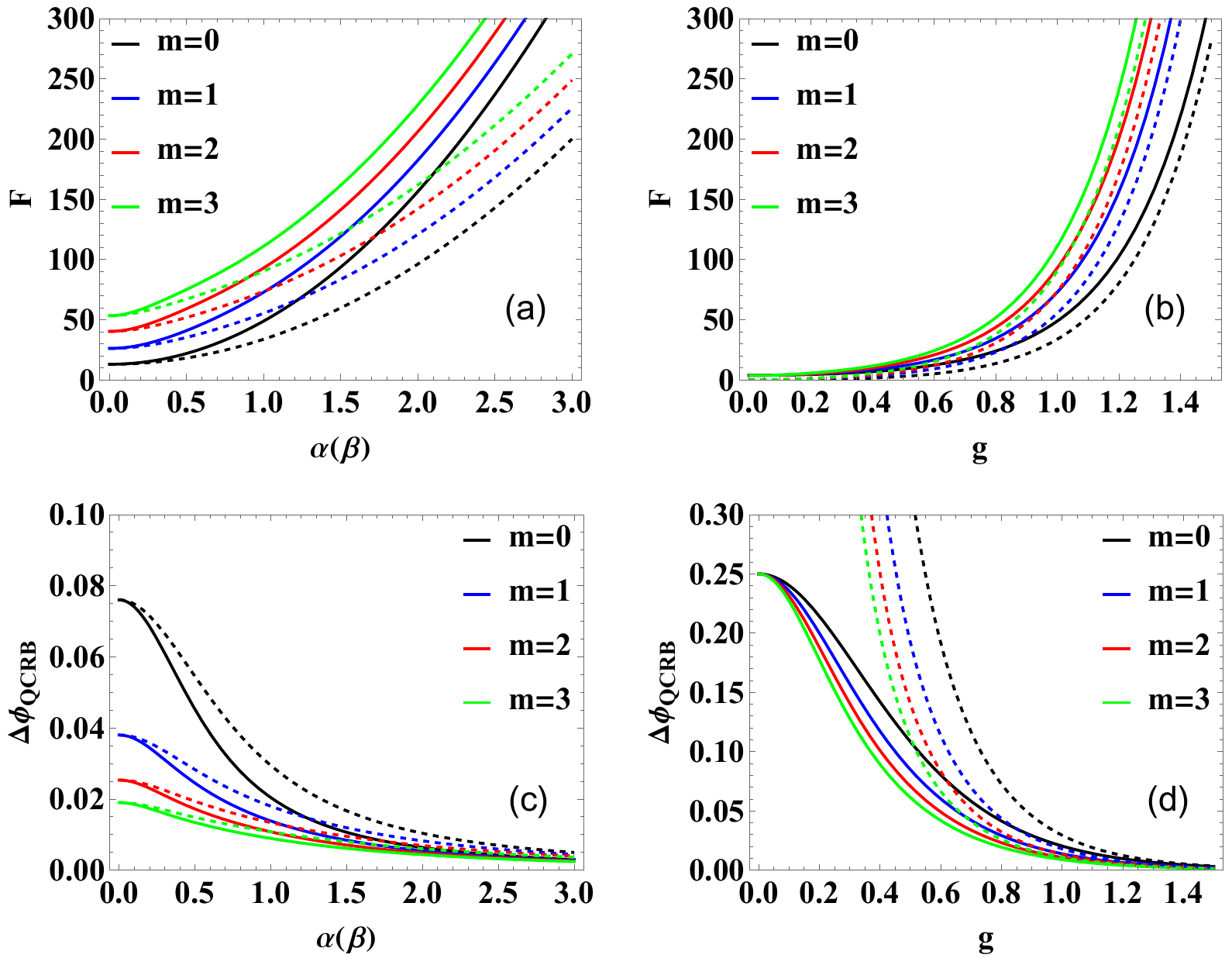}%
}
\caption{The QFI as a function of (a) $\protect \alpha (\protect \beta )$,
with $\protect \phi =\protect \pi /2$, $\protect \tau =0.5$, and $g=1$; (b) $g$%
, with $\protect \phi =\protect \pi /2$, $\protect \tau =0.5$, and $\protect%
\alpha (\protect \beta )=1$. The QCRB as a function of (c) $\protect \alpha (%
\protect \beta )$, with $\protect \phi =\protect \pi /2$, $\protect \tau =0.5$,
and $g=1$; (d) $g$, with $\protect \phi =\protect \pi /2$, $\protect \tau =0.5$%
, and $\protect \alpha (\protect \beta )=1$. The solid and dashed lines
represent Schemes A and B, respectively.}
\end{figure*}

In Fig. 8, we present the QFI and the QCRB as functions of the coherent
amplitude $\alpha (\beta )$ and gain factor $g$. QFI appears in Fig. 8(a)
and (b), and QCRB in Fig. 8(c) and (d). It can be concluded that: (i) QFI
and QCRB improve significantly with\ increasing\ $\alpha (\beta )$ and $g$.
This is because higher $\alpha (\beta )$ increases the input photon number,
while higher $g$ enhances OPA nonlinearity; (ii) Our schemes enhance QFI and
QCRB due to the non-Gaussian nature, with further improvement as
photon-subtracted number $m$ increases; (iii) The improvement first
increases and then decreases with increasing $\alpha (\beta )$ and $g$. This
is because when quantum resources are limited, PS has a stronger effect. As $%
\alpha (\beta )$ and $g$ grow, more photons and stronger nonlinear effects
improve phase sensitivity but reduce PS impact;\ (iv) Scheme A outperforms
Scheme B. However, when $\alpha (\beta )\rightarrow 0$, both states approach
vacuum, leading to similar performance. At $g$ $\rightarrow 0$, Scheme A
performs better because more input light passes through the phase shifter,
carrying more phase information.

In the traditional scheme, QFI is independent of both the phase and the
vBS's transmittance. In contrast, in our scheme, the PS at the output
undergoes a sequential interaction with the phase shifter and the vBS,
carrying phase information and being modulated by transmittance. Thus, we
plot QFI as a function of phase $\phi $ and transmittance $\tau $ under two
schemes with PS, and show how the QCRB varies with these parameters in Fig.
9.

\begin{figure*}[tbp]
\label{Figure9} {\centering \includegraphics[width=1.4\columnwidth]{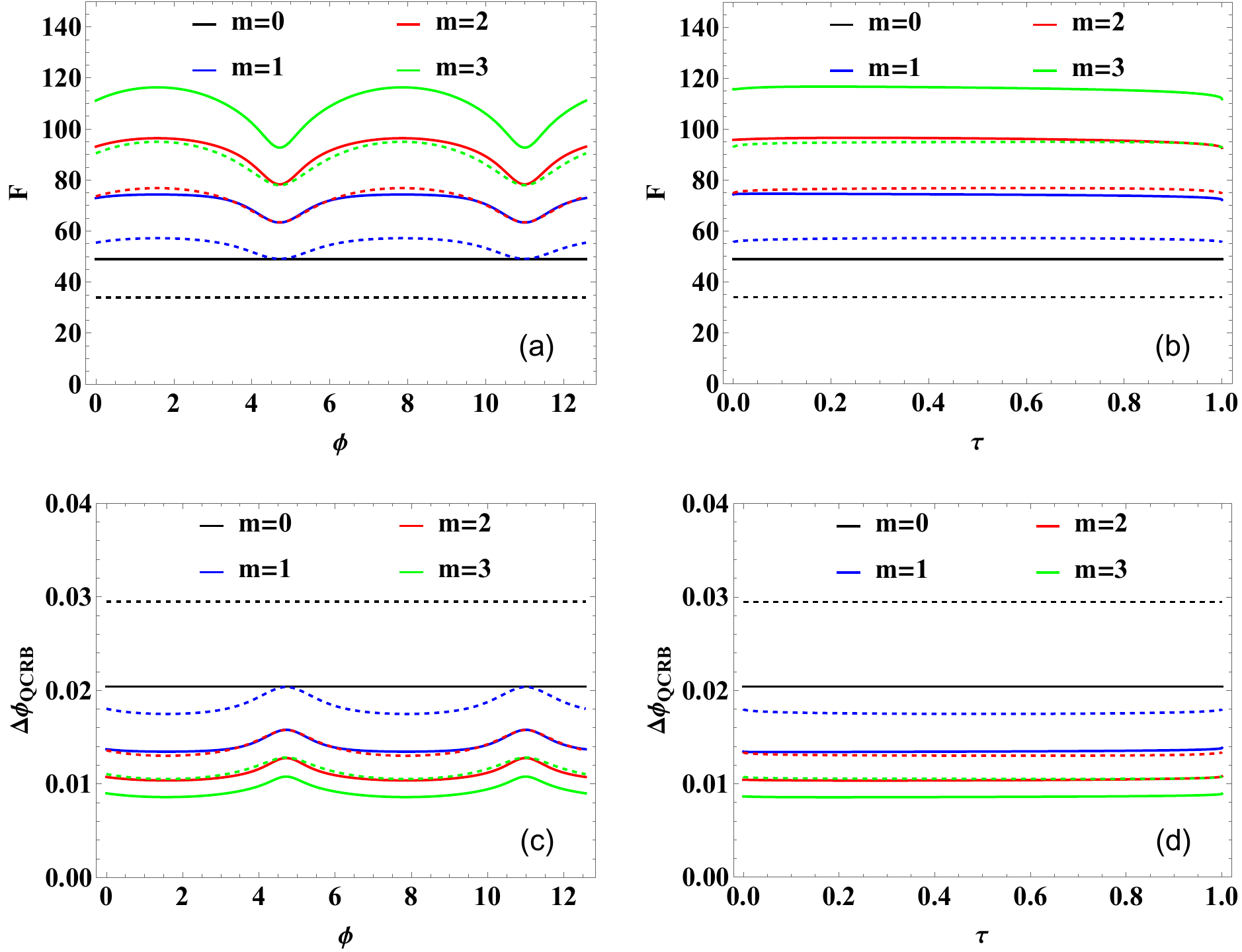}%
}
\caption{The QFI as a function of (a) $\protect \phi $,\ with $\protect \tau %
=0.5$, $\protect \alpha (\protect \beta )=1$,\ and $g=1$; (b) $\protect \tau $,
with\ $\protect \phi =\protect \pi /2$,\ $\protect \alpha (\protect \beta )=1$,\
and $g=1$. The QCRB as a function of (c) $\protect \phi $,\ with $\protect%
\tau =0.5$, $\protect \alpha (\protect \beta )=1$,\ and $g=1$; (d) $\protect%
\tau $, with\ $\protect \phi =\protect \pi /2$,\ $\protect \alpha (\protect%
\beta )=1$,\ and $g=1$. The solid and dashed lines represent Schemes A and
B, respectively.}
\end{figure*}

As shown in Fig. 9, for the PS scheme, (i) QFI varies periodically with the
phase $\phi $ with a period of $2\pi $, reaching maximum values at $\pi
/2+2n\pi $ and minimum values at $3\pi /2+2n\pi $, where $n$ is an integer
(see Fig. 9(a)). This indicates that QFI can be effectively optimized by
adjusting $\phi $; (ii) Scheme\ A achieves higher QFI than Scheme B, with
more pronounced phase fluctuations, which become more significant as $m$
increases; (iii) however, QFI is insensitive to changes in $\tau $ (see Fig.
9(b)), which differs from the behavior of phase sensitivity. Therefore, $%
\tau $ is fixed at $0.5$ in all subsequent QFI and QCRB analyses. Moreover,
QCRB shows similar results as these parameters change, as shown in Fig. 9(c)
and (d).

\subsection{QFI in the lossy case}

This subsection extends the analysis to QFI in the presence of photon loss.
For simplicity, we focus on the photon loss occurring before and after the $%
a $-mode phase shifter, which is modeled by a virtual BS with transmittance $%
\eta $, as shown in Fig. 10. Following the approach proposed by Escher et
al., the QFI under loss conditions can be obtained \cite{50}. For details,
please refer to Supplement 1, S3. Similar to the ideal case, one can compute
the QCRB as $\Delta \phi _{QCRB_{L}}=1/\sqrt{vF_{L}}$,\ and for simplicity
we take $v=1$. Next, we analyze the degradation of the QFI and QCRB caused
by photon loss.

\begin{figure}[tbh]
\label{Figure10} \centering \includegraphics[width=0.85%
\columnwidth]{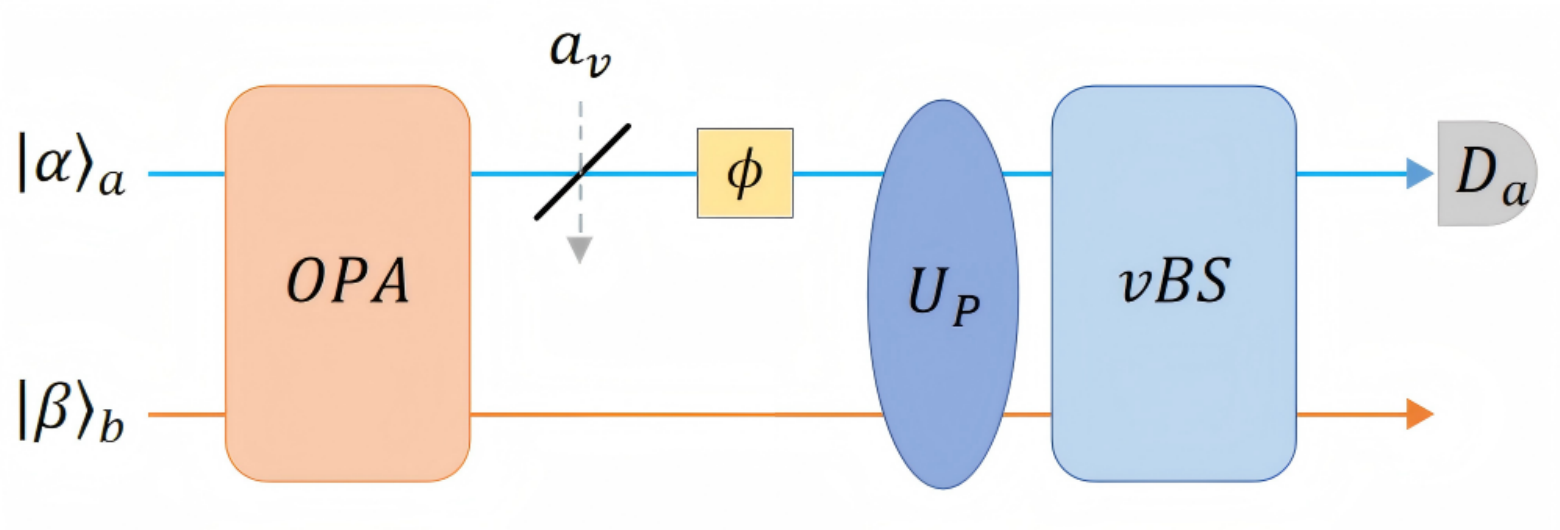}
\caption{Schematic diagram of the photon loss on mode $a$, where $U_{P}$
represents a non-local operation, and $U_{P}=B^{\dag }a^{m}B$.}
\end{figure}

In Fig. 11, we plot the QFI and QCRB as functions of the transmittance $\eta 
$. It can be observed that: (i) The QFI and QCRB decrease as $\eta $
decreases, because photon loss reduces the system's average photon number;
(ii) QFI and QCRB are improve by PS, with greater enhancement at higher $m$;
(iii) Scheme A outperforms Scheme B. Notably, subtracting two photons in
Scheme A nearly matches subtracting three in Scheme B, demonstrating
superior performance under photon loss. 
\begin{figure}[tph]
\label{Figure11} \centering \includegraphics[width=0.85%
\columnwidth]{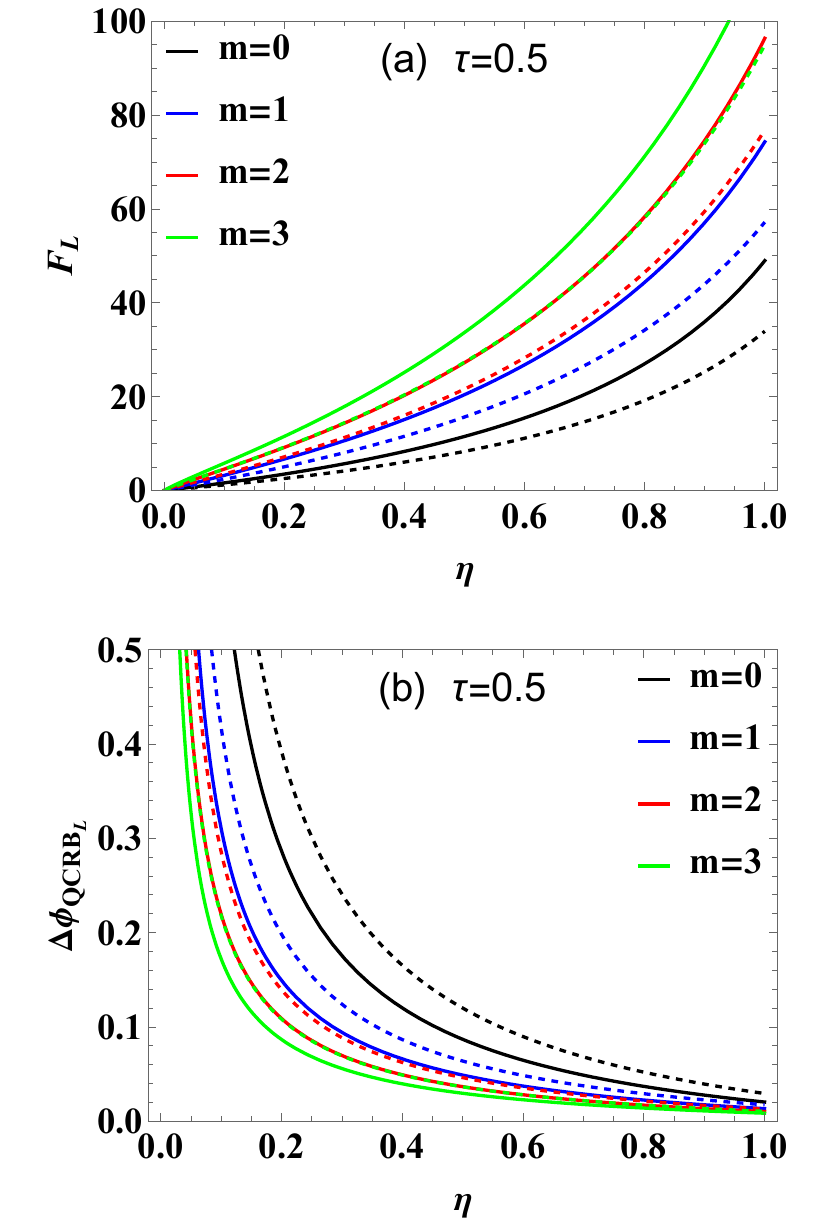} \ 
\caption{(a) The $F_{L}$ as a function of transmittance $\protect \eta $,
with\ $\protect \phi =\protect \pi /2$,\ $\protect \tau =0.5$,\ $\protect \alpha %
\left( \protect \beta \right) =1$. (b) The QCRB$_{L}$ as a function of
transmittance $\protect \eta $, with\ $\protect \phi =\protect \pi /2$,\ $%
\protect \tau =0.5$,\ $\protect \alpha \left( \protect \beta \right) =1$. The
solid and dashed lines represent Schemes A and B, respectively.}
\label{11}
\end{figure}

\subsection{Phase sensitivity compared with theoretical limits}

Furthermore, we compare phase sensitivity with key theoretical limits: the
SQL ($1/\sqrt{N_{T}}$), the HL ($1/N_{T}$), and the QCRB, where $N_{T}$ is
the average photon number inside the interferometer. We use the equivalent
model in Fig. 10 to calculate $N_{T}$, which is located after the non-local
operation $U_{P}$ and before the equivalent vBS. The expression for $N_{T}$
is as follows

\begin{equation}
N_{T}=\left. _{int}\left \langle \Psi \right \vert \left( a^{\dagger
}a+b^{\dagger }b\right) \left \vert \Psi \right \rangle _{int}\right. ,
\label{c8}
\end{equation}%
where $\left \vert \Psi \right \rangle _{int}=N_{3}U_{p}U_{\phi
}B_{T}S_{g}\left \vert \alpha \right \rangle _{a}\left \vert \beta
\right
\rangle _{b}\left \vert 0\right \rangle _{a_{v}}$. Further
computational details are provided in Supplement 1, S4.

\begin{figure}[tph]
\label{Figure12} \centering \includegraphics[width=0.85%
\columnwidth]{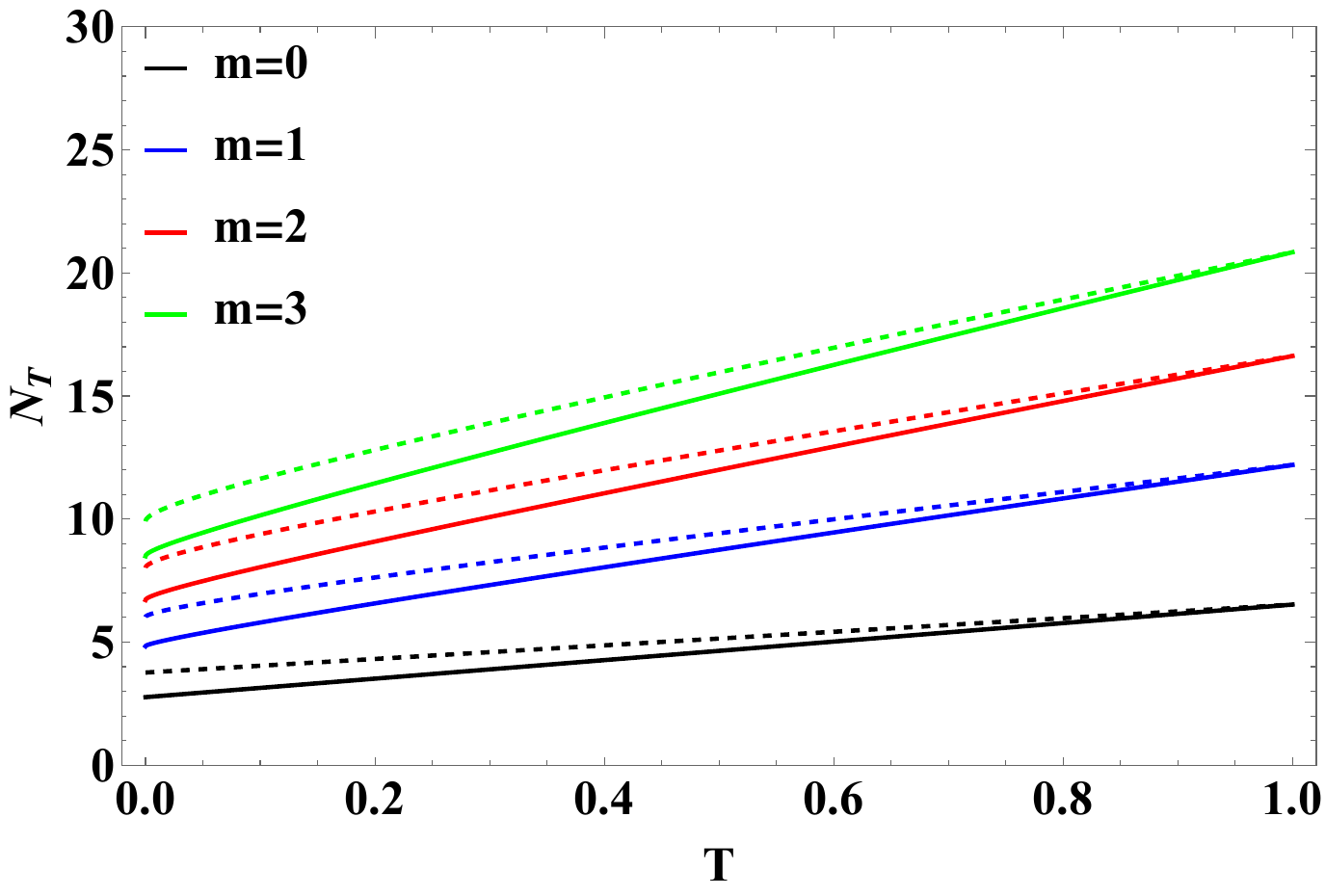} \ 
\caption{The average photon number $N$ as a function of $T$, with $\protect%
\phi =\protect \pi /2$, $\protect \tau =0.5$, $\protect \alpha \left( \protect%
\beta \right) =1$, and $g=1$. The solid and dashed lines represent Schemes A
and B, respectively.}
\label{12}
\end{figure}

In Fig. 12, we exhibit that for the two input schemes, the internal average
photon number changes with the transmittance $T$ of the vBS, which models
the photon loss in mode $a$. The following observations can be made: (i) In
the ideal case $(T=1)$, both schemes with PS produce the same average photon
number, which increases as $m$ increases; (ii) As $T$ decreases, the photon
number drops, with Scheme A showing a larger decline than Scheme B. This is
because loss is limited to mode $a$, and Scheme A mainly introduces photons
through this mode; (iii) At $T=0$, photon numbers remain positive in both
schemes, because photons in mode $a$ are completely lost, while photons in
mode $b$ pass through the interferometer. These results confirm that the
input mode and photon loss distribution strongly affect internal photon
number distribution. 
\begin{figure*}[tph]
\label{Figure13} \centering \includegraphics[width=1.4%
\columnwidth]{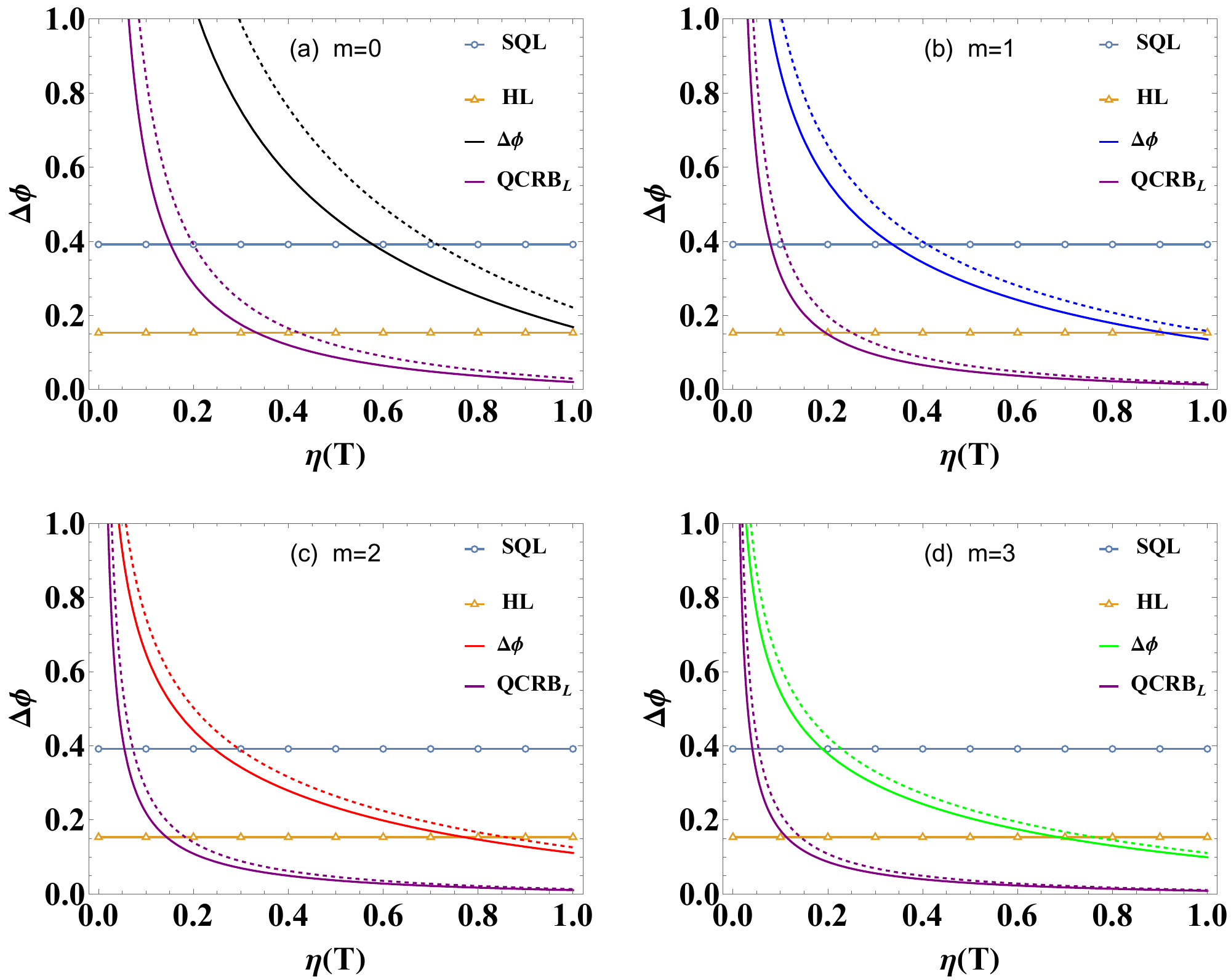}
\caption{Comparison of the phase sensitivity with SQL and HL. In the figure,
the blue circle represents SQL and the yellow triangle represents HL. The
solid and dashed lines represent Schemes A and B, respectively.}
\end{figure*}

For simplicity in analysis, we set $T=\eta $. Fig. 13 shows the phase
sensitivity and QCRB as a function of the transmittance $T$($\eta $) for two
input schemes, compared with the SQL and HL.\ Solid and dashed lines
represent Schemes A and B, respectively.\ The main findings are as follows:
(i)\ PS significantly improves system robustness against photon loss. For
instance, when $m=0$, Scheme A and B fail to surpass the SQL at
approximately 40\% and 30\% loss, respectively; however, when $m=3$, the
loss tolerance thresholds of both increase to about 80\%. Notably, our
schemes at $m=3$ can surpass the HL even under 20\% loss, while the original
scheme without PS cannot; (ii) As $m$ increases, robustness improves further
and phase sensitivity approaches the QCRB more closely; (iii) Scheme A
outperforms Scheme B in phase sensitivity and robustness,\ with its
advantage increasing as loss rises. Overall, our scheme demonstrates clear
advantages under lossy conditions.

\section{Conclusion}

In this paper, we propose a phase\ estimation protocol using PS at the
output port of an OPA-BS hybrid interferometer with a coherent state and a
vacuum state as inputs.\ Two input configurations are considered: Scheme A
(mode $a$ injection) and Scheme B (mode $b$ injection), with homodyne
detection on mode $a$. We evaluate phase sensitivity and QFI under both
ideal and lossy conditions. Results show that PS significantly improves both
metrics, enhancing system robustness. The enhancement increases with the PS
order $m$. Scheme A outperforms Scheme B in sensitivity and robustness,
indicating a synergistic effect between the $a$-mode coherent state and PS,
which benefits performance in noisy environments. Comparisons with
theoretical limits show that our scheme at$\ m=3$ can surpass the HL even
under $20\%$ loss, and the loss tolerance threshold for exceeding the SQL is
increased from the original 30\%-40\% to $\sim $80\%.

Moreover, for phase sensitivity, the optimal transmittance $\tau $ of vBS is
approximately 0.5 under ideal conditions. However, under photon loss, the
optimal $\tau $ increases with the level of loss, indicating that adjusting
vBS's transmittance to optimize resource allocation\ can effectively
compensate for loss and enhance robustness. Additionally, QFI is affected by
the phase and transmittance $\tau $ due to output-port PS. Research shows
that QFI varies periodically with phase and reaches its maximum at $\pi /2$,
but it is largely insensitive to $\tau $. In summary, combining PS with a
vBS can effectively improve performance under lossy conditions. This work
offers a new approach to high-precision phase measurement using non-Gaussian
operations with simple input resources.

\begin{acknowledgments}
\bigskip This work is supported by the National Natural Science Foundation
of China (Grants No.12564049 and No. 12104195) and the Jiangxi Provincial
Natural Science Foundation (Grants No. 20242BAB26009 and 20232BAB211033),
as well as the Jiangxi Provincial Key Laboratory of Advanced Electronic Materials and
Devices (Grant No. 2024SSY03011), Science and Technology Project of 
Jiangxi Provincial Department of Education (Grant No. GJJ2404102)
and Jiangxi Civil-Military Integration Research Institute (Grant No. 2024JXRH0Y07).
\end{acknowledgments}

\end{document}


\title{Phase estimation via multi-photon subtraction at the output of the
hybrid interferometer}
\author{}

\section{The phase sensitivity}

\bigskip \label{S1} In this appendix, we provide the calculation formulas of
the phase sensitivity with PS based on homodyne detection. In our scheme,
the output state is given by%
\begin{equation}
\left \vert \Psi \right \rangle _{out}=N_{0}a^{m}B_{v}U_{\phi
}B_{T}S_{g}\left \vert \Psi \right \rangle _{in},  \label{a1}
\end{equation}%
where, $N_{0}$ is the normalization constant, the input state of the
expanded space is $\left \vert \Psi \right \rangle _{in}=\left \vert \alpha
\right \rangle _{a}\otimes \left \vert \beta \right \rangle _{b}\otimes
\left \vert 0\right \rangle _{a_{v}}$.\ When $\alpha =1,\beta =0$, it
corresponds to input Scheme A; when $\alpha =0,\beta =1$, it corresponds to
input Scheme B.

To simplify, we calculate the general formula of the operator expectation
value $\left. _{out}\left \langle \Psi \right \vert a^{\dagger
k}a^{l}\left
\vert \Psi \right \rangle _{out}\right. $. In order to derive
the general formula, we introduce a formula, i.e.,%
\begin{equation}
a^{\dagger k}a^{l}=\frac{\partial ^{k+l}}{\partial t^{k}\partial s^{l}}%
e^{ta^{\dagger }}e^{sa}|_{t=s=0},  \label{a2}
\end{equation}%
and apply the transformation relations, i.e.,%
\begin{eqnarray}
&&S_{g}^{\dag }B_{T}^{\dag }U_{\phi }^{\dag }B_{v}^{\dag }aB_{v}U_{\phi
}B_{T}S_{g}  \notag \\
&=&(ae^{i\phi }\sqrt{\tau T}\cosh g-ia^{\dagger }e^{i\theta }\sqrt{1-\tau }%
\sinh g  \notag \\
&&+ib\sqrt{1-\tau }\cosh g-b^{\dagger }e^{i\phi }e^{i\theta }\sqrt{\tau T}%
\sinh g+a_{v}\sqrt{\tau }\sqrt{1-T}e^{i\phi }).  \label{a3}
\end{eqnarray}%
Here, $k$ and $l$ are positive integers, and $s$ and $t$ are the
differential variables. After the differentiation, all these differential
variables take zero.

Thus, the mathematical analytic form of $\left. _{out}\left \langle \Psi
\right \vert a^{\dagger k}a^{l}\left \vert \Psi \right \rangle _{out}\right. 
$ can be expressed as 
\begin{eqnarray}
\left. _{out}\left \langle \Psi \right \vert a^{\dagger k}a^{l}\left \vert
\Psi \right \rangle _{out}\right. &=&N_{0}^{2}\left. _{in}\left \langle \Psi
\right \vert S_{g}^{\dag }B_{T}^{\dag }U_{\phi }^{\dag }B_{v}^{\dag
}a^{\dagger k}a^{l}B_{v}U_{\phi }B_{T}S_{g}\left \vert \Psi \right \rangle
_{in}\right.  \notag \\
&=&N_{0}^{2}G_{k,l}e^{Z_{0}},  \label{a4}
\end{eqnarray}%
where%
\begin{equation}
G_{k,l}=\frac{\partial ^{k+l}}{\partial t^{k}\partial s^{l}}\left \{ \cdot
\right \} |_{t=s=0}  \label{a5}
\end{equation}%
and%
\begin{equation}
Z_{0}=s^{2}X_{1}+t^{2}X_{1}^{\ast }+s\alpha X_{2}+t\alpha X_{2}^{\ast
}+s\beta X_{3}+t\beta X_{3}^{\ast }+st\left( 1-\tau +\tau T\right) \sinh
^{2}g,  \label{a6}
\end{equation}%
with%
\begin{eqnarray}
X_{1} &=&-ie^{i\phi }e^{i\theta }\sqrt{\tau T}\sqrt{1-\tau }\sinh g\cosh g, 
\notag \\
X_{2} &=&e^{i\phi }\sqrt{\tau T}\cosh g-ie^{i\theta }\sqrt{1-\tau }\sinh g, 
\notag \\
X_{3} &=&i\sqrt{1-\tau }\cosh g-e^{i\phi }e^{i\theta }\sqrt{\tau T}\sinh g, 
\notag \\
X_{4} &=&\left( 1-\tau +\tau T\right) \sinh ^{2}g.  \label{a7}
\end{eqnarray}

According to Eq. \ref{a4}, the normalization constant can be expressed as%
\begin{equation}
N_{0}=\left( G_{m,m}e^{Z_{0}}\right) ^{-\frac{1}{2}},  \label{a8}
\end{equation}%
and the phase sensitivity of our scheme is%
\begin{equation}
\Delta \phi =\frac{\sqrt{\left \langle X^{2}\right \rangle -\left \langle
X\right \rangle ^{2}}}{|\partial _{\phi }\left \langle X\right \rangle |},
\label{a9}
\end{equation}%
where%
\begin{eqnarray}
\left \langle X\right \rangle &=&\left. _{out}\left \langle \Psi \right
\vert (a+a^{\dagger })\left \vert \Psi \right \rangle _{out}\right. /\sqrt{2}%
,  \notag \\
&=&N_{0}^{2}(G_{m,m+1}+G_{m+1,m})e^{Z_{0}}/\sqrt{2},  \label{a10}
\end{eqnarray}%
\begin{eqnarray}
\left \langle X^{2}\right \rangle &=&\left. _{out}\left \langle \Psi \right
\vert (a^{2}+a^{\dagger 2}+2a^{\dagger }a+1)\left \vert \Psi \right \rangle
_{out}\right. /2  \notag \\
&=&[N_{0}^{2}(G_{m,m+2}+G_{m+2,m}+2G_{m+1,m+1})e^{Z_{0}}+1]/2.  \label{a11}
\end{eqnarray}%
\  \ 

When $T=1$, Eq. \ref{a9} corresponds to the phase sensitivity under ideal
conditions.

\section{QFI under ideal condition}

To calculate the ideal QFI, we introduce an equivalent model shown in Fig.
1. The ideal QFI can be given by%
\begin{equation}
F=4\left[ \left \langle \psi _{\phi }^{\prime }|\psi _{\phi }^{\prime
}\right \rangle -\left \vert \left \langle \psi _{\phi }^{\prime }|\psi
_{\phi }\right \rangle \right \vert ^{2}\right] ,  \label{b1}
\end{equation}%
where $\left \vert \psi _{\phi }\right \rangle =N_{1}U_{P}U_{\phi
}S_{g}\left \vert \alpha \right \rangle _{a}\left \vert \beta \right \rangle
_{b} $ represents the quantum state after the phase shifter, with the
non-local operation $U_{P}=B_{v}^{\dag }a^{m}B_{v}$, and derived $%
\left
\vert \psi _{\phi }^{\prime }\right \rangle =\partial \left \vert
\Psi _{\phi }\right \rangle /\partial \phi $. Here, the normalization
constant can be expressed\ as 
\begin{equation}
N_{1}=\left( D_{m,0,0,0,0}e^{Z_{1}}\right) ^{-\frac{1}{2}},  \label{b2}
\end{equation}%
and 
\begin{equation}
\left \langle \psi _{\phi }^{^{\prime }}\right \vert \left \vert \psi _{\phi
}^{^{\prime }}\right \rangle =N_{1}^{2}D_{m,1,1,1,1}e^{Z_{1}}-iN_{1}\frac{%
\partial N_{1}}{\partial \phi }D_{m,1,1,0,0}e^{Z_{1}}+iN_{1}\frac{\partial
N_{1}}{\partial \phi }D_{m,0,0,1,1}e^{Z_{1}}+\left( \frac{\partial N_{1}}{%
\partial \phi }\right) ^{2}D_{m,0,0,0,0}e^{Z_{1}},  \label{b3}
\end{equation}%
\begin{equation}
\left \langle \psi _{\phi }^{\prime }\right \vert \left \vert \psi _{\phi
}\right \rangle =-iN_{1}^{2}D_{m,1,1,0,0}e^{Z_{1}}+N_{1}\frac{\partial N_{1}%
}{\partial \phi }D_{m,0,0,0,0}e^{Z_{1}},  \label{b4}
\end{equation}%
\begin{equation}
\left \langle \psi _{\phi }\right \vert \left \vert \psi _{\phi }^{\prime
}\right \rangle =iN_{1}^{2}D_{m,0,0,1,1}e^{Z_{1}}+N_{1}\frac{\partial N_{1}}{%
\partial \phi }D_{m,0,0,0,0}e^{Z_{1}},  \label{b5}
\end{equation}%
where%
\begin{equation}
D_{m,x_{1},y_{1},x_{2},y_{2}}=\frac{\partial ^{2m}}{\partial t^{m}\partial
s^{m}}\frac{\partial ^{x_{1}+y_{1}+x_{2}+y_{2}}}{\partial c^{x_{1}}\partial
d^{y_{1}}\partial p^{x_{2}}\partial h^{y_{2}}}\left \{ \cdot \right \}
|_{t=s=c=d=p=h=0}  \label{b6}
\end{equation}%
and%
\begin{eqnarray}
Z_{1} &=&(Z_{0}+psy_{1}y_{3}^{\ast }+tpy_{1}y_{4}+dy_{1}^{\ast }\left(
py_{1}+ty_{3}+sy_{4}^{\ast }\right) +hy_{2}^{\ast }\left(
py_{2}+sy_{5}^{\ast }\right)  \notag \\
&&+ty_{6}\left( hy_{2}^{\ast }+sy_{6}^{\ast }\right) +cy_{2}\left(
dy_{2}^{\ast }+hy_{2}^{\ast }+ty_{5}+sy_{6}^{\ast }\right)  \notag \\
&&+\left( cy_{1}+py_{1}+dy_{1}^{\ast }+hy_{1}^{\ast }\right) \alpha +\left(
cy_{2}+py_{2}+dy_{2}^{\ast }+hy_{2}^{\ast }\right) \beta ,  \label{b7}
\end{eqnarray}%
\begin{equation}
Z_{0}=s^{2}X_{1}+t^{2}X_{1}^{\ast }+s\alpha X_{2}+t\alpha X_{2}^{\ast
}+s\beta X_{3}+t\beta X_{3}^{\ast }+st\left( 1-\tau \right) \sinh
^{2}g,\left( T=1\right)  \label{b8}
\end{equation}%
with%
\begin{eqnarray}
y_{1} &=&e^{-i\phi }\cosh g,  \notag \\
y_{2} &=&-e^{-i\phi }e^{-i\theta }\sinh g,  \notag \\
y_{3} &=&e^{-i\phi }\sqrt{\tau }\cosh g,  \notag \\
y_{4} &=&ie^{-i\theta }\sqrt{1-\tau }\sinh g,  \notag \\
y_{5} &=&-i\sqrt{1-\tau }\cosh g,  \notag \\
y_{6} &=&-e^{-i\phi }e^{-i\theta }\sqrt{\tau }\sinh g.  \label{b9}
\end{eqnarray}%
Here, $m$, $x_{1}$, $y_{1}$, $x_{2}$ and $y_{2}$ are positive integers, and $%
s$, $t$, $c$, $d$, $p$ and $h$ are the differential variables. After the
differentiation, all these differential variables take zero.

\section{QFI in the presence of photon loss}

\begin{figure}[tbh]
\label{Figure1} \centering \includegraphics[width=0.85%
\columnwidth]{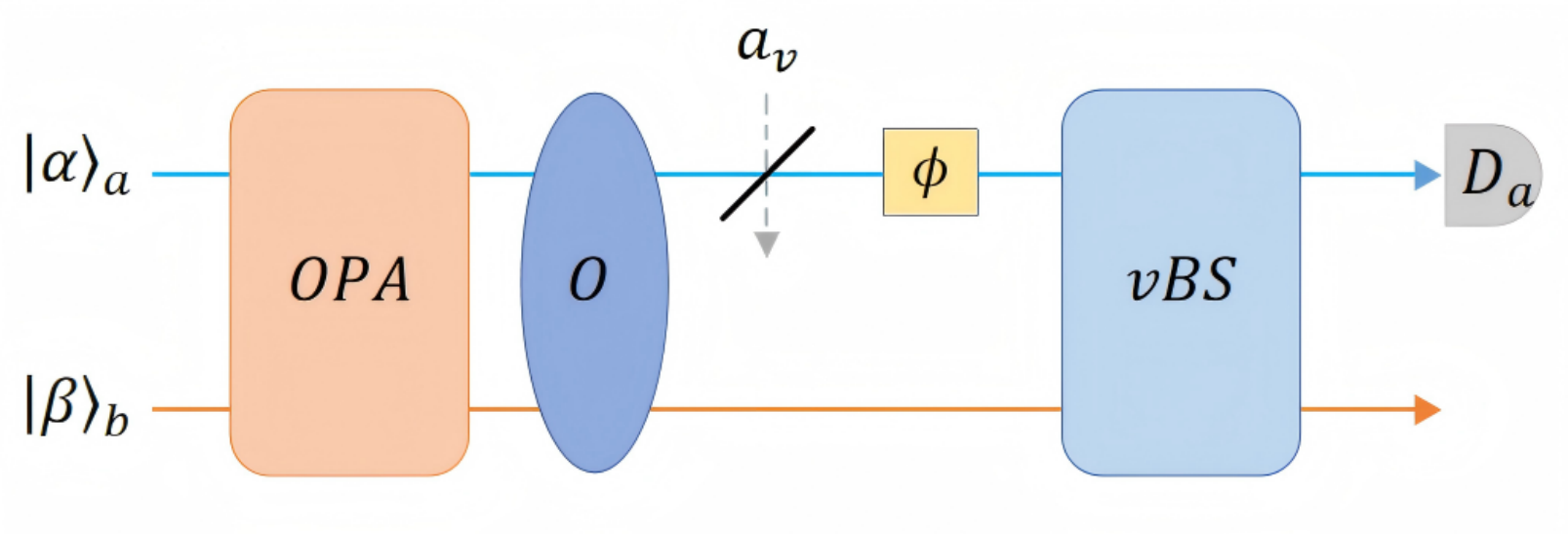}
\caption{Schematic diagram of the photon losses on mode $a$, where $O$
represents a non-local operation, and $O=\frac{\partial ^{m}}{\partial s^{m}}%
\exp \left[ s\left( ae^{i\protect \phi }\protect \sqrt{\protect \tau \protect%
\eta }+ib\protect \sqrt{1-\protect \tau }\right) \right] |_{s=0}$}
\end{figure}

Next, we derive the QFI under photon losses, as shown in Fig. 1. For the
hybrid interferometer, the QFI quantifies the amount of information acquired
before the vBS, so we focus on the quantum state prior to it. The initial
probe state of the interferometer system $S$ is denoted as $\left \vert \psi
\right \rangle _{S}=S_{g}\left \vert \alpha \right \rangle _{a}\otimes
\left
\vert \beta \right \rangle _{b}.$ Due to photon losses, phase
encoding process is no longer unitary. Thus, we reformulate\ the problem as
parameter estimation under unitary evolution $U_{S+E}\left( \phi \right) $
in an extended system $S+E$, where the state $\left \vert \psi
\right
\rangle _{S+E}$ evolves unitarily and is given by

\begin{align}
\left \vert \psi \right \rangle _{S+E}& =N_{2}U_{P}U_{S+E}\left( \phi
\right) \left \vert \psi \right \rangle _{S}\left \vert 0\right \rangle _{E}
\notag \\
& =N_{2}U_{S+E}\left( \phi \right) O\left \vert \psi \right \rangle
_{S}\left \vert 0\right \rangle _{E}  \notag \\
& =N_{2}\overset{\infty }{\underset{l=0}{\sum }}\Pi _{l}\left( \phi \right)
O\left \vert \psi \right \rangle _{S}\left \vert l\right \rangle _{E}  \notag
\\
& =\overset{\infty }{\underset{l=0}{\sum }}\Pi _{l}\left( \phi \right) \left
\vert \Psi \right \rangle \left \vert l\right \rangle _{E}.  \label{c1}
\end{align}%
Here, $\left \vert 0\right \rangle _{E}$ refers to the initial state of the
environment subject to photon losses, $\left \vert l\right \rangle _{E}$
denotes an orthogonal basis relative to $\left \vert 0\right \rangle _{E}$,
and $\left \vert \Psi \right \rangle =N_{2}O\left \vert \psi \right \rangle
_{S}$, where $N_{2}$ is the normalization factor and $O$ represents the
equivalent operator that transforms the non-local operation $U_{P}$ into a
form applied prior to the photon losses. The expression for $\hat{O}$ is
given by $O=\frac{\partial ^{m}}{\partial s^{m}}\exp \left[ s\left(
ae^{i\phi }\sqrt{\tau \eta }+ib\sqrt{1-\tau }\right) \right] |_{s=0}$.
Additionally, $\Pi _{l}\left( \phi \right) $ is the Kraus operator used to
describe photon losses, and its form is given by \cite{1}

\begin{equation}
\Pi _{l}\left( \phi \right) =\sqrt{\frac{\left( 1-\eta \right) ^{l}}{l!}}%
e^{i\phi \left( n_{a}-\lambda l\right) }\eta ^{\frac{n_{a}}{2}}a^{l},
\label{c2}
\end{equation}%
where, $e^{i\phi n_{a}}$\ is the operator of phase shifter. The parameter $%
\eta $ represents the transmittance of the virtual BS that simulating photon
losses.\ When $\eta =0$ or $1$, it corresponds to complete absorption and
lossless conditions, respectively. We introduce the parameter $\lambda $, $%
\lambda =0$ or $-1$ corresponding to photon losses after or before the phase
shift, respectively.

In this extended system $S+E$, the QFI under loss conditions is given by 
\begin{equation}
F_{L}=C_{Q}\left[ \left \vert \psi \right \rangle _{S},\Pi _{l}\left( \phi
\right) \right] _{\min },  \label{c3}
\end{equation}%
where, $\left \vert \psi \right \rangle _{S}$ is the initial state of the
detection system S, and $C_{Q}$ denotes the QFI of the extended system.%
\begin{equation}
C_{Q}\left[ \left \vert \psi \right \rangle _{S},\Pi _{l}\left( \phi \right) %
\right] =4\left[ \left. _{S+E}\left \langle \psi ^{\prime }|\psi ^{\prime
}\right \rangle _{S+E}\right. -\left \vert _{S+E}\left \langle \psi ^{\prime
}|\psi \right \rangle _{S+E}\right \vert ^{2}\right] .  \label{c4}
\end{equation}

Based on Eqs. (\ref{c1}) and (\ref{c2}), $C_{Q}\left[ \left \vert \psi
\right \rangle _{S},\Pi _{l}\left( \phi \right) \right] $ can be expressed as

\begin{equation}
C_{Q}\left[ \left \vert \psi \right \rangle _{S},\Pi _{l}\left( \phi \right) %
\right] =4\left[ \left \langle h_{1}\right \rangle -\left \vert \left
\langle h_{2}\right \rangle \right \vert ^{2}\right] ,  \label{c5}
\end{equation}%
with $\left \langle \cdot \right \rangle =\left. _{S}\left \langle \psi
\right \vert \cdot \left \vert \psi \right \rangle _{S}\right. $, and

\begin{align}
h_{1}& =\overset{\infty }{\underset{l=0}{\sum }}\frac{dN_{2}O_{l}^{\dagger
}\Pi ^{\dagger }\left( \phi \right) }{d\phi }\frac{dN_{2}\Pi _{l}\left( \phi
\right) \hat{O}}{d\phi }  \notag \\
& =\frac{dN_{2}O^{\dagger }}{d\phi }\frac{dN_{2}O}{d\phi }+N_{2}O^{\dagger
}H_{1}N_{2}O+\frac{dN_{2}O^{\dagger }}{d\phi }iH_{2}^{\dagger
}N_{2}O+N_{2}O^{\dagger }\left( -iH_{2}\right) \frac{dN_{2}O}{d\phi },
\label{c6}
\end{align}%
\begin{equation}
h_{2}=i\left. \overset{\infty }{\underset{l=0}{\sum }}\frac{dN_{2}O^{\dagger
}\Pi _{l}^{\dagger }\left( \phi \right) }{d\phi }\Pi _{l}\left( \phi \right)
N_{2}O\right. =i\frac{dN_{2}O^{\dagger }}{d\phi }N_{2}O+N_{2}O^{\dagger
}H_{2}N_{2}O,  \label{c7}
\end{equation}%
where%
\begin{align}
H_{1}& =\overset{\infty }{\underset{l=0}{\sum }}\frac{d\Pi _{l}^{\dagger
}\left( \phi \right) }{d\phi }\frac{d\Pi _{l}\left( \phi \right) }{d\phi },
\label{c8} \\
H_{2}& =i\overset{\infty }{\underset{l=0}{\sum }}\frac{d\Pi _{l}^{\dagger
}\left( \phi \right) }{d\phi }\Pi _{l}\left( \phi \right) .  \label{c9}
\end{align}

Based on Eqs. (\ref{c2}) and (\ref{c5})-(\ref{c9}), $C_{Q}\left[ \left \vert
\psi \right \rangle _{S},\Pi _{l}\left( \phi \right) \right] $ is given by

\begin{eqnarray}
C_{Q} &=&4\left[ \left \langle h_{1}\right \rangle -\left \vert \left
\langle h_{2}\right \rangle \right \vert ^{2}\right]  \notag \\
&=&4[\left \langle \tilde{\Psi}\right \vert \left \vert \tilde{\Psi}\right
\rangle -\left \vert \left \langle \tilde{\Psi}\right \vert \left \vert \Psi
\right \rangle \right \vert ^{2}+\left \langle \Psi \right \vert H_{1}\left
\vert \Psi \right \rangle -\left \vert \left \langle \Psi \right \vert
H_{2}\left \vert \Psi \right \rangle \right \vert ^{2}  \notag \\
&&+\left \langle \tilde{\Psi}\right \vert iH_{2}^{\dagger }\left \vert \Psi
\right \rangle -\left \langle \Psi \right \vert iH_{2}\left \vert \tilde{\Psi%
}\right \rangle +i\left \langle \Psi \right \vert H_{2}\left \vert \Psi
\right \rangle \left \langle \Psi \right \vert \left \vert \tilde{\Psi}%
\right \rangle -i\left \langle \tilde{\Psi}\right \vert \left \vert \Psi
\right \rangle \left \langle \Psi \right \vert H_{2}^{\dagger }\left \vert
\Psi \right \rangle ],  \label{c10}
\end{eqnarray}%
where $\left \vert \tilde{\Psi}\right \rangle =\partial \left \vert \Psi
\right \rangle /\partial \phi $, and $H_{1}$, $H_{2}$ being expressed by 
\cite{2}%
\begin{eqnarray}
H_{1} &=&\left[ 1-\left( 1+\lambda \right) \left( 1-\eta \right) \right]
^{2}n_{a}^{2}+\left( 1+\lambda \right) ^{2}\eta \left( 1-\eta \right) n_{a},
\label{c11} \\
H_{2} &=&\left[ 1-\left( 1+\lambda \right) \left( 1-\eta \right) \right]
n_{a}.  \label{c12}
\end{eqnarray}

To find the minimum $C_{Q}$, we set $\partial C_{Q}\left[ \left \vert \psi
\right \rangle _{S},\Pi _{l}\left( \phi \right) \right] /\partial \lambda =0.$
By optimizing $\lambda $ to obtain $C_{Q\min }$, and the QFI can be
expressed as%
\begin{eqnarray}
F_{Q} &=&4\left[ \left \langle h_{1}\right \rangle -\left \vert \left \langle
h_{2}\right \rangle \right \vert ^{2}\right] _{\min }  \notag \\
&=&4[N_{2}^{2}\left. _{S}\left \langle \psi \right \vert \frac{dO^{\dag }}{%
d\phi }\frac{dO}{d\phi }\left \vert \psi \right \rangle _{S}\right.
-N_{2}^{2}\left. _{S}\left \langle \psi \right \vert \frac{dO^{\dag }}{d\phi }%
O\left \vert \psi \right \rangle _{S}\right. N_{2}^{2}\left. _{S}\left \langle
\psi \right \vert O^{\dag }\frac{dO}{d\phi }\left \vert \psi \right \rangle
_{S}\right. +N_{2}^{2}\left. _{S}\left \langle \psi \right \vert O^{\dag
}n_{a}^{2}O\left \vert \psi \right \rangle _{S}\right. -(N_{2}^{2}\left.
_{S}\left \langle \psi \right \vert O^{\dag }n_{a}O\left \vert \psi
\right \rangle _{S}\right. )^{2}  \notag \\
&&+i(N_{2}^{2}\left. _{S}\left \langle \psi \right \vert \frac{dO^{\dag }}{%
d\phi }n_{a}O\left \vert \psi \right \rangle _{S}\right. -N_{2}^{2}\left.
_{S}\left \langle \psi \right \vert O^{\dag }n_{a}\frac{dO}{d\phi }\left \vert
\psi \right \rangle _{S}\right. )+iN_{2}^{2}\left. _{S}\left \langle \psi
\right \vert O^{\dag }n_{a}O\left \vert \psi \right \rangle _{S}\right.
(N_{2}^{2}\left. _{S}\left \langle \psi \right \vert O^{\dag }\frac{dO}{d\phi }%
\left \vert \psi \right \rangle _{S}\right. -N_{2}^{2}\left. _{S}\left \langle
\psi \right \vert \frac{dO^{\dag }}{d\phi }O\left \vert \psi \right \rangle
_{S}\right. )]  \notag \\
&&-\frac{\left( 1-\eta \right) }{\left( 1-\eta \right) [N_{2}^{2}\left.
_{S}\left \langle \psi \right \vert O^{\dag }n_{a}^{2}O\left \vert \psi
\right \rangle _{S}\right. -(N_{2}^{2}\left. _{S}\left \langle \psi
\right \vert O^{\dag }n_{a}O\left \vert \psi \right \rangle _{S}\right.
)^{2}]+\eta N_{2}^{2}\left. _{S}\left \langle \psi \right \vert O^{\dag
}n_{a}O\left \vert \psi \right \rangle _{S}\right. }  \notag \\
&&\times \{2[N_{2}^{2}\left. _{S}\left \langle \psi \right \vert O^{\dag
}n_{a}^{2}O\left \vert \psi \right \rangle _{S}\right. -(N_{2}^{2}\left.
_{S}\left \langle \psi \right \vert O^{\dag }n_{a}O\left \vert \psi
\right \rangle _{S}\right. )^{2}]+i(N_{2}^{2}\left. _{S}\left \langle \psi
\right \vert \frac{dO^{\dag }}{d\phi }n_{a}O\left \vert \psi \right \rangle
_{S}\right. -N_{2}^{2}\left. _{S}\left \langle \psi \right \vert O^{\dag }n_{a}%
\frac{dO}{d\phi }\left \vert \psi \right \rangle _{S}\right. )  \notag \\
&&+iN_{2}^{2}\left. _{S}\left \langle \psi \right \vert O^{\dag
}n_{a}O\left \vert \psi \right \rangle _{S}\right. (N_{2}^{2}\left.
_{S}\left \langle \psi \right \vert O^{\dag }\frac{dO}{d\phi }\left \vert \psi
\right \rangle _{S}\right. -N_{2}^{2}\left. _{S}\left \langle \psi \right \vert 
\frac{dO^{\dag }}{d\phi }O\left \vert \psi \right \rangle _{S}\right. )\}^{2},
\label{c13}
\end{eqnarray}%
where

\begin{equation}
N_{2}=\left( P_{m,0,0}e^{Z_{2}}\right) ^{-\frac{1}{2}},  \label{c14}
\end{equation}%
\begin{equation}
\left. _{S}\left \langle \psi \right \vert \frac{dO^{\dagger }}{d\phi }\frac{%
dO}{d\phi }\left \vert \psi \right \rangle _{S}\right. =\tau \eta
P_{m,1,1}\left( ste^{Z_{2}}\right) |_{t=s=0},  \label{c15}
\end{equation}%
\begin{equation}
\left. _{S}\left \langle \psi \right \vert \hat{O}^{\dagger }\frac{d\hat{O}}{%
d\phi }\left \vert \psi \right \rangle _{S}\right. =ie^{i\phi }\sqrt{\tau
\eta }P_{m,0,1}\left( se^{Z_{2}}\right) |_{t=s=0},  \label{c16}
\end{equation}%
\begin{equation}
\left. _{S}\left \langle \psi \right \vert \frac{d\hat{O}^{\dagger }}{d\phi }%
n_{a}\hat{O}\left \vert \psi \right \rangle _{S}\right. =-ie^{-i\phi }\sqrt{%
\tau \eta }P_{m,2,1}\left( te^{Z_{2}}\right) |_{t=s=0},  \label{c17}
\end{equation}%
\begin{equation}
\left. _{S}\left \langle \psi \right \vert O^{\dagger }n_{a}O\left \vert
\psi \right \rangle _{S}\right. =P_{m,1,1}\left( e^{Z_{2}}\right) |_{t=s=0},
\label{c18}
\end{equation}%
\begin{equation}
\left. _{S}\left \langle \psi \right \vert O^{\dagger }n_{a}^{2}O\left \vert
\psi \right \rangle _{S}\right. =\left( P_{m,2,2}+P_{m,1,1}\right)
e^{Z_{2}}|_{t=s=0},  \label{c19}
\end{equation}%
and%
\begin{equation}
P_{m,k,l}=\frac{\partial ^{2m}}{\partial t^{m}\partial s^{m}}\frac{\partial
^{k+l}}{\partial x^{k}\partial y^{l}}\left \{ \cdot \right \} |_{t=s=x=y=0},
\label{c20}
\end{equation}%
\begin{equation}
Z_{2}=e^{Z_{0}+sxX_{5}+tyX_{5}^{\ast }+syX_{6}+txX_{6}^{\ast }+xy\sinh
^{2}g+x\alpha \cosh g+y\alpha \cosh g-y\beta e^{i\theta }\sinh g-x\beta
e^{-i\theta }\sinh g},  \label{c21}
\end{equation}%
as well as%
\begin{equation}
Z_{0}=s^{2}X_{1}+t^{2}X_{1}^{\ast }+s\alpha X_{2}+t\alpha X_{2}^{\ast
}+s\beta X_{3}+t\beta X_{3}^{\ast }+stX_{4},  \label{c22}
\end{equation}

\begin{eqnarray}
X_{1} &=&-ie^{i\phi }e^{i\theta }\sqrt{\tau \eta }\sqrt{1-\tau }\sinh g\cosh
g,  \notag \\
X_{2} &=&e^{i\phi }\sqrt{\tau \eta }\cosh g-ie^{i\theta }\sqrt{1-\tau }\sinh
g,  \notag \\
X_{3} &=&i\sqrt{1-\tau }\cosh g-e^{i\phi }e^{i\theta }\sqrt{\tau \eta }\sinh
g,  \notag \\
X_{4} &=&\left( 1-\tau +\tau \eta \right) \sinh ^{2}g,  \notag \\
X_{5} &=&e^{i\phi }\sqrt{\tau \eta }\sinh ^{2}g,  \notag \\
X_{6} &=&-ie^{i\theta }\sqrt{1-\tau }\sinh g\cosh g.  \label{c23}
\end{eqnarray}

Here, $m$, $k$ and $l$ are positive integers, and $s$, $t$, $x$ and $y$ are
the differential variables. After the differentiation, all these
differential variables take zero.

\section{Internal mean photon number}

In the equivalent model, the expression of the average internal photon
number is given by%
\begin{eqnarray}
N &=&\left. _{int}\left \langle \Psi \right \vert \left( a^{\dagger
}a+b^{\dagger }b\right) \left \vert \Psi \right \rangle _{int}\right.  \notag
\\
&=&N_{3}^{2}(Q_{m,1,1,0,0}+Q_{m,0,0,1,1})e^{Z_{3}},  \label{d1}
\end{eqnarray}%
where $\left \vert \Psi \right \rangle _{int}=N_{3}U_{p}U_{\phi
}B_{T}S_{g}\left \vert \alpha \right \rangle _{a}\left \vert \beta
\right
\rangle _{b}\left \vert 0\right \rangle _{a_{v}}$, with $%
U_{p}=B_{v}^{\dag }a^{m}B_{v}$, and%
\begin{eqnarray}
Z_{3} &=&Z_{0}+s\left( f_{1}\lambda _{1}+f_{2}\lambda _{2}+f_{3}\lambda
_{3}+f_{4}\lambda _{4}\right) +t\left( f_{1}^{\ast }\lambda _{2}+f_{2}^{\ast
}\lambda _{1}+f_{3}^{\ast }\lambda _{4}+f_{4}^{\ast }\lambda _{3}\right) 
\notag \\
&&+f_{5}\lambda _{1}\lambda _{2}+f_{6}\lambda _{2}\lambda _{4}+f_{6}^{\ast
}\lambda _{1}\lambda _{3}+f_{7}\lambda _{3}\lambda _{4}  \notag \\
&&+\left( f_{8}^{\ast }\lambda _{1}+f_{8}\lambda _{2}+f_{9}^{\ast }\lambda
_{3}+f_{9}\lambda _{4}\right) \alpha +\left( f_{10}^{\ast }\lambda
_{1}+f_{10}\lambda _{2}+\lambda _{3}f_{11}+\lambda _{4}f_{11}\right) \beta ,
\label{d2}
\end{eqnarray}%
\begin{equation}
Z_{0}=X_{1}s^{2}+X_{1}^{\ast }t^{2}+s\alpha X_{2}+t\alpha X_{2}^{\ast
}+s\beta X_{3}+t\beta X_{3}^{\ast }+stX_{4},  \label{d3}
\end{equation}%
\begin{equation}
Q_{m,l_{1},l_{2},l_{3},l_{4}}=\frac{\partial ^{2m}}{\partial t^{m}\partial
s^{m}}\frac{\partial ^{l_{1}+l_{2}+l_{3}+l_{4}}}{\partial \lambda
_{1}^{l_{1}}\partial \lambda _{2}^{l_{2}}\partial \lambda
_{4}^{l_{3}}\partial \lambda _{2}^{l_{4}}}\left \{ \cdot \right \}
|_{t=s=\lambda _{1}=\lambda _{2}=\lambda _{3}=\lambda _{4}=0},  \label{d4}
\end{equation}%
as well as%
\begin{equation}
N_{3}=\left( Q_{m,0,0,0,0}e^{Z_{3}}\right) ^{-\frac{1}{2}}.  \label{d5}
\end{equation}%
\begin{eqnarray}
X_{1} &=&-ie^{i\phi }e^{i\theta }\sqrt{\tau \eta }\sqrt{1-\tau }\sinh g\cosh
g,  \notag \\
X_{2} &=&e^{i\phi }\sqrt{\tau \eta }\cosh g-ie^{i\theta }\sqrt{1-\tau }\sinh
g,  \notag \\
X_{3} &=&i\sqrt{1-\tau }\cosh g-e^{i\phi }e^{i\theta }\sqrt{\tau \eta }\sinh
g,  \notag \\
X_{4} &=&\left( 1-\tau +\tau \eta \right) \sinh ^{2}g,  \label{d6}
\end{eqnarray}%
\begin{eqnarray}
f_{1} &=&\sqrt{\tau }T\sinh ^{2}g,  \notag \\
f_{2} &=&-ie^{i\phi }e^{i\theta }\sqrt{T}\sqrt{1-\tau }\sinh g\cosh g, 
\notag \\
f_{3} &=&i\sqrt{1-\tau }\sinh ^{2}g,  \notag \\
f_{4} &=&-e^{i\phi }e^{i\theta }\sqrt{\tau T}\sinh g\cosh g,  \notag \\
f_{5} &=&T\sinh ^{2}g,  \notag \\
f_{6} &=&-e^{i\phi }e^{i\theta }\sqrt{T}\sinh g\cosh g,  \notag \\
f_{7} &=&\sinh ^{2}g,  \notag \\
f_{8} &=&e^{i\phi }\sqrt{T}\cosh g,  \notag \\
f_{9} &=&-e^{i\theta }\sinh g,  \notag \\
f_{10} &=&-e^{i\phi }e^{i\theta }\sqrt{T}\sinh g.  \label{d7}
\end{eqnarray}%
\newline

\bibliographystyle{plain}
\bibliography{sample}